\documentclass[11pt]{iopart}
\pdfoutput=1
\usepackage{amstext,amssymb,xspace}
\usepackage{graphicx}
\usepackage{xcolor}
\usepackage{cite}
\newcommand{\grad}{{ \nabla}}
\newcommand{\avg}[1]{\left\langle #1 \right\rangle}	% for average
\newcommand{\abs}[1]{\left| #1 \right|}	% for absolute
\newcommand{\dt}{{\rm d}t}
\newcommand{\dx}{{\rm d}x}

\renewcommand{\d}[2]{\frac{{\rm d} #1}{{\rm d} #2}} 	% for derivatives

\usepackage{tikz}
\renewcommand{\vec}[1]{\mathbf{#1}}					% for bold font vector 
\newcommand{\eqref}[1]{(\ref{#1})}
\newcommand{\bpmatrix}{\left(\begin{array}{cc}}
\newcommand{\epmatrix}{\end{array}\right)}

\begin{document}
 \title{Lattice Models of Nonequilibrium Bacterial Dynamics}
\author{A.G. Thompson, J. Tailleur, M.E. Cates and R.A. Blythe}
\address{SUPA, School of Physics and Astronomy, University of
Edinburgh, Kings Buildings, Mayfield Road, Edinburgh EH9 3JZ, UK}

\begin{abstract}
 We study a model of self propelled particles exhibiting run-and-tumble dynamics on lattice. This non-Brownian diffusion is characterised by a random walk with a finite persistence length between changes of direction, {and is inspired by the motion of bacteria such as {\em E. coli}}. By defining a class of models with multiple species of particle and transmutation between species we can recreate such dynamics. These models admit exact analytical results whilst also forming a counterpart to previous continuum models of run-and-tumble dynamics. We solve the externally driven non-interacting and zero-range versions of the model exactly and utilise a field theoretic approach to derive the continuum fluctuating hydrodynamics for more general interactions. We make contact with prior approaches to run-and-tumble dynamics off lattice and determine the steady state and linear stability for a class of crowding interactions, where the jump rate decreases as density increases. {In addition to its interest from the perspective of nonequilibrium statistical mechanics, this lattice model constitutes and efficient tool to simulate a class of interacting run-and-tumble models relevant to bacterial motion, so long as certain conditions (that we derive) are met.}
\end{abstract}

\maketitle

\section{Introduction}
\label{sec:Introduction}

The last decade has seen a growing number of studies of biological systems conducted by physicists. Much of this work relies on tools not traditionally found in biology. For instance, recent methods of nonlinear optics have made possible observation of biological systems on much smaller scales than was previously possible~\cite{Westphal2008}. In other cases biological systems have helped shed light on questions of fundamental importance in theoretical physics. Studies of bird flocks showed that it may be possible to observe long range order in systems with a continuous symmetry if detailed balance is broken~\cite{Toner1995} (in equilibrium this is forbidden by the Mermin-Wagner theorem~\cite{Mermin1966}).

Bacteria provide one example of a biological system which is of great interest to the study of nonequilibrium physics. Some species of bacteria, such as \emph{Escherichia coli}, swim by means of a series of relatively straight runs interspersed with {short} periods of chaotic motion, known as tumbles, during which their orientation changes at random and they experience little net displacement~\cite{Berg2004}. This can be seen as a type of non-Brownian diffusion where the steady state probability distribution will not be Boltzmann. 

Following experiments to determine their behaviour by Berg and others~\cite{Berg1972,Berg1973,Berg2003,berg00,Larsen1974,Macnab1977,Turner2000} since the 1970s, much is known about the dynamics and behaviour of individual bacteria. Less, however, is known about their collective behaviour and it is here that statistical mechanics can play a useful role. Schnitzer et al. analysed various strategies bacteria may employ, e.g. changing their speed or tumble rate, and studied the differences these make to the steady state distribution~\cite{Schnitzer1993}. {These analyses, which were based on off-lattice descriptions of the particle dynamics, were more recently extended to allow for various types of interaction between bacteria~\cite{Tailleur2008}, external force fields~\cite{Tailleur2009}, hydrodynamic interactions\cite{Nash2010}, and coupling to logistic population growth~\cite{Cates2010}. Intriguingly it was shown that a dependence of motility parameters such as swim-speed on density was alone enough to cause phase separation~\cite{Tailleur2008}, in contrast to detailed-balance systems where many-body effects on particle mobility cancel exactly in Boltzmann steady states.

In this work we construct a lattice model whose non-interacting continuum limit recovers the off lattice equations previously derived for noninteracting run-and-tumble particles~\cite{Schnitzer1993,Tailleur2009}. We will also address interactions in the form of density dependent motility parameters, and thereby connect also with the off lattice approach of ~\cite{Tailleur2008}.  
One motivation for this approach is as follows. Although real bacterial systems are off lattice, both the microscopic run-and-tumble equations, and the diffusive continuum equations found by coarse graining these, are difficult to simulate efficiently, particularly once interactions are included. On lattice the local density of particles is easy to determine and so density-dependent interactions are easy to include at relatively low cost computationally. Lattice simulations are also easier to extend to higher dimensions as we will consider towards the end of this paper. For all these reasons, creation of a robust and accurate lattice representation of bacterial motility is a worthwhile goal, even from a purely phenomenological standpoint, whereby the purpose of a model is to provide a fairly direct explanation for results seen in experiment.

A second motivation is more fundamental. Models of non-equilibrium statistical mechanics can be broadly split into two categories: one phenomenological as just described, the other comprising simple models which allow the study of basic concepts and facilitate a more detailed understanding of the nature of non-equilibrium physics. In the latter category we can think of lattice transport models such as the exclusion~\cite{Blythe2007} or zero-range~\cite{Evans2005} processes, for which some exact analytical results can be found, as well as methods to characterise fluctuations and large deviations in non-equilibrium states~\cite{Derrida2007}. Our model falls squarely into this category of simple theoretical models, and indeed extends some of these examples; we investigate both a zero-range process and a partially excluding version of our run-and-tumble system. As a microscopic lattice model we can, under certain conditions, calculate exact steady states and understand precisely how changes in the underlying dynamics affect the probability distributions. More generally we can always write an exact master equation for probability of a configuration and utilise a variety of field theoretic representations to derive the large scale fluctuating hydrodynamics, to attempt to map the system to a free energy and to determine the steady state behaviour and dynamic stability.

As already explained, in addition to being an interesting toy model that we can use to understand driven non-equilibrium physics, our model also forms a direct lattice counterpart to some well established phenomenological continuum models of bacterial dynamics. 
It is unusual for a model to allow exact computations while credibly describing the real behaviour of a physical system, and this is one of the reasons why we consider this model to be of interest.}

In section~\ref{sec:Presentation} we introduce our model of run-and-tumble bacteria on a lattice. In  section~\ref{sec:ExactSolutions} we consider those cases in which we can find exact solutions to the steady state behaviour. After first examining in section~\ref{sec:NonIntSystems} the non-interacting case, which for inhomogeneous and anisotropic jump and tumble rates can still admit non-trivial solutions, in section~\ref{sec:ZeroRange} we include a zero-range interaction and derive sufficient conditions on the rates for a factorised steady state to exist. 

In section~\ref{sec:FHIB} we consider more general interactions, where the jump rates can depend on the full configuration of the system, and derive a continuum fluctuating hydrodynamics using a field theoretic approach. We first set up the theoretical apparatus and re-derive the results for non-interacting systems within this framework; we then calculate the field equations for the more generic interacting system. In section~\ref{sec:CrowdingInt} we consider one specific type of interaction and investigate the effect of crowding. We derive a free-energy-like functional from which we can determine the steady state probability distribution and compare the effects of different methods to coarse grain the local density in the interaction terms. Finally we extend our results to higher dimensions more applicable to real systems in section~\ref{sec:TwoDimensions}.

\section{The Model}
\label{sec:Presentation}

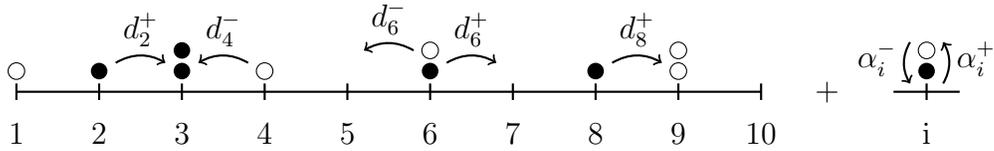
\begin{figure}
\centering
  \begin{tikzpicture}[scale=1.1]
    \foreach \y in {3}{
      \draw[thick] (1,\y) -- (10,\y);
      \foreach \x in {1,...,10}{
	\draw[thick] (\x,\y-.1) -- (\x,\y+.1);
      }
    }
     \foreach \y in {2.5}{
      \foreach \x in {1,...,10}{
      \draw (\x,\y) node {$\x$};
      }
    }

      \foreach \x in {1,4,9}{ 
	\draw (\x,3.25) circle (.1);
      }
   
      \foreach \x in {2,3,6,8}{
         \fill (\x,3.25) circle (.1);
      }

      \draw (9,3.5) circle (.1);

      \draw (6,3.5) circle (.1);

      \fill (3,3.5) circle (.1);

    \foreach \x in {2,8}{
      \draw[thick,->] (\x+.2,3.35) to [bend left] (\x+.8,3.35);
      \draw (\x+.5,3.75) node {$d^+_{\x}$};
    }

      \draw[thick,->] (6+.2,3.35) to [bend left] (6+.8,3.35);
      \draw (6+.5,3.75) node {$d^+_{6}$};

      \draw[thick,->] (6-.2,3.5) to [bend right] (6-.8,3.5);
      \draw (6-.5,3.9) node {$d^-_{6}$};

      \foreach \x in {4}{
	\draw[thick,->] (\x-.2,3.35) to [bend right] (\x-.8,3.35);
	\draw (\x-.5,3.75) node {$d^-_{\x}$};
      }

	\draw (10.8,3) node {+};
	\draw (12,2.5) node {i};
	\draw[thick] (11.6,3) -- (12.4,3);
	\draw[thick] (12,2.9) -- (12,3.1);
      
	\draw (12,3.5) circle (.1);
	\fill (12,3.25) circle (.1);

      \draw[thick,->] (12-.2,3.6) to [bend right] (12-.2,3.1) ;
      \draw (12-.6,3.4) node {$\alpha^-_{i}$};

      \draw[thick,->] (12+.2,3.1) to [bend right] (12+.2,3.6);
      \draw (12+.6,3.4) node {$\alpha^+_{i}$};

    \end{tikzpicture}
  \caption{Presentation of the model. Filled circles represent right moving particles while unfilled circles denote left moving particles. Some of the possible transitions are illustrated on the figure.}
  \label{fig:model}
\end{figure}

To model the finite persistence length in run-and-tumble dynamics on lattice particles jump repeatedly in the same direction, $\vec{u}$, with rate $d(\vec{u})$ and change direction - tumble - with rate $\alpha(\vec{u})$. In 1d, this means particles can be either right-going or left-going. Right-going
particles jump with rate $d^+_i$ from site $i$ to site $i+1$ and
tumble with rate $\alpha_i^+$. After a tumble, they become left-going
particles with probability $1/2$. The corresponding rates for
left-going particles are called $d^-_i$ and $\alpha^-_i$, see figure~\ref{fig:model}. Throughout this paper we assume tumbles to be instantaneous. In reality the duration of tumbles is typically of the order of one tenth of the duration of runs~\cite{Berg2004}. There may be situations, however, where time spent tumbling may increase, and where the finite duration of a tumble may have an effect on the dynamics and steady state of the system. We leave a consideration of this case to future work.

\section{Exactly Solvable Cases}
\label{sec:ExactSolutions}
The jump and tumble rates defined in section~\ref{sec:Presentation} may, in general, depend on the occupation numbers of any lattices sites. Through this we may introduce interactions to our system and consider the general case of $N$ interacting bacteria.  Before we deal with such complex cases in section~\ref{sec:FHIB}, however, let us look at the interesting limiting cases that allow for exact computation of the steady state. In particular we start by considering first the non-interacting case and investigating the effect of position and direction dependence on the rates in section~\ref{sec:NonIntSystems}, before then turning to zero-range processes in section~\ref{sec:ZeroRange}. 
\subsection{Non-Interacting Particles}
\label{sec:NonIntSystems}
For non-interacting particles, one can always handle the single particle
case first and compute the average occupancy $\rho_i^\pm$ of left ($-$) or right ($+$) moving particles on site $i$. As shown below, the steady-state distribution of $n$ particles is then given by a product
on each site of a multinomial distribution of parameters $\rho_i^\pm$.

Let us start with the single particle process and call $P(k,+)$ and
$P(k,-)$ the probability to find a bacterium at site $k$ going to the
right and to the left, respectively. The master equation reads
\begin{eqnarray}
    \fl\partial_t P(k,+)&=&d_{k-1}^+ P(k-1,+) -d_k^+ P(k,+) +\frac{\alpha_k^-}2 P(k,-) - \frac{\alpha_k^+}2 P(k,+)\\
    \fl \partial_t P(k,-)&=&d^-_{k+1} P(k+1,-) -d^-_k P(k,-) -\frac{\alpha^-_k}2 P(k,-) + \frac{\alpha^+_k}2 P(k,+).
  \label{eqn:master}
\end{eqnarray}

\subsubsection{Exact Solution of the Steady State}
For the case of non-interacting particles, where all interesting and non-trivial effects come from heterogeneities or anisotropies in the jump and tumble rates, we can calculate the steady state probability distributions exactly. We start from the continuity equation
\begin{equation}
\d{}{t}\big[P(i,+)+P(i,-)\big]=J_{i-1,i}-J_{i,i+1},
\label{eq:Continuity}
\end{equation}
where $J_{i,i+1}$ is the net probability flux between sites $i$ and $i+1$, that is
\begin{equation}
 J_{i,i+1}=d_i^+\,P(i,+)-d_{i+1}^-\,P(i+1,-).
\label{eq:LatticeCurrentDef}
\end{equation}
As we are only after the steady state distribution we take the time derivative on the left hand side of equation~\eqref{eq:Continuity} to be zero, so all currents $J_{i,i+1}$ are constant and equal to some $J$ fixed by the boundary conditions. 

The master equation for the density of left-moving particles at site $i$ reads
\begin{equation}
 0 = d_{i+1}^-\,P(i+1,-) - d_i^-\,P(i,-) +\frac{\alpha_i^+}{2}P(i,+) - \frac{\alpha_i^-}{2}P(i,-).
\label{eq:NonIntME}
\end{equation}
Using \eqref{eq:LatticeCurrentDef} to eliminate the first term on the right hand side of equation~\eqref{eq:NonIntME}, which is the only term to depend on site $i+1$, we can establish a relationship between the number of left and right moving particles on site $i$ at steady state:
\begin{equation}
 P(i,+)\left(d_i^++\frac{\alpha_i^+}{2}\right)-P(i,-)\left(d_i^-+\frac{\alpha_i^-}{2}\right)=J.
\label{eq:LatticeCurrentTwo}
\end{equation}
Equation~\eqref{eq:LatticeCurrentTwo}, along with equation~\eqref{eq:LatticeCurrentDef}, leads to the recursion relation for right moving particles
\begin{equation}
P(i+1,+)=\frac{d_i^+}{d_{i+1}^-}\frac{2\,d_{i+1}^-+\alpha_{i+1}^-}{2\,d_{i+1}^++\alpha_{i+1}^+}P(i,+)-J\frac{\alpha_{i+1}^-}{d^-_{i+1}(\alpha_{i+1}^++2\,d^+_{i+1})},
\label{eq:LatticeRecursion}
\end{equation}
which can be solved to yield
\begin{eqnarray}
 P(i,+)&=&\prod_{j=1}^{i-1}\left(\frac{d_j^+}{d_{j+1}^-}\frac{2\,d_{j+1}^-+\alpha_{j+1}^-}{2\,d_{j+1}^++\alpha_{j+1}^+}\right)\Bigg(P(1,+)\nonumber\label{eq:LatticeNIProbA} \\
&-&J\sum_{k=1}^{i-1}\frac{\alpha_{k+1}^-}{d^-_{k+1}\,(2\,d^+_{k+1}+\alpha_{k+1}^+)\prod_{m=1}^{k}\left(\frac{d_j^+}{d_{j+1}^-}\frac{2\,d_{j+1}^-+\alpha_{j+1}^-}{2\,d_{j+1}^++\alpha_{j+1}^+}\right)}\Bigg)\\
P(i,-)&=&\frac{\left(2\,d_i^++\alpha_i^+\right)P(i,+)-J}{2\,d_i^-+\alpha_i^-}.
\label{eq:LatticeNIProbB}
\end{eqnarray}
The probability to find a particle at any position and in either state can then be calculated by noting that the total distribution must be normalised, i.e. $\sum_i\left[P(i,+)+P(i,-)\right]=1$ and that $J$ is imposed by the boundary conditions. For example, closed boundaries require that $J=0$, while for periodic boundaries we have the additional constraint that $P(L+1,\pm)=P(1,\pm)$. Note that the probability densities for left and right moving particles do not have to be the same, and, in general, will not be.

The probability of a given configuration $P(\left\lbrace n_i^+,n_i^-\right\rbrace)$, is given by
\begin{equation}
 P\left(\left\lbrace n_i^+,n_i^-\right\rbrace\right) = N! \prod_i \frac{P(i,+)^{n_i^+}P(i,-)^{n_i^-}}{n_i^+!\,n_i^-!},
\end{equation}
where $N$ is the total number of particles in the system. If we call the average number of right or left going particles on a site $\rho_i^\pm=N\,P(i,\pm)$, then in the limit where $N\rightarrow\infty$ and $L\rightarrow\infty$, so that $P(i,\pm)\rightarrow0$, but $\rho_i^\pm$ remains finite, the probability of a configuration is given by
\begin{equation}
 P\left(\left\lbrace n_i^+,n_i^-\right\rbrace\right) = \prod_i \frac{(\rho_i^+)^{n_i^+}\,\exp(-\rho_i^+)}{n_i^+!}\frac{(\rho_i^-)^{n_i^-}\,\exp(-\rho_i^-)}{n_i^-!}.
\end{equation}

\subsubsection{Continuous limit}
Since we ultimately want to compare the run-and-tumble on lattice with its off lattice counterpart, let us first take the continuum limit of the master equation. Explicitly introducing the lattice spacing $a$ and defining
$x_k= ka$, the master equation \eqref{eqn:master} reads
\begin{eqnarray}
\fl    \partial_t P(x_k,+)&=&d^+(x_{k}-a) P(x_k-a,+) -d^+(x_k) P(x_k,+)\nonumber\\
\fl &+&\frac{\alpha^-(x_k)}2 P(x_k,-) - \frac{\alpha^+(x_k)}2 P(x_k,+)\\
\fl     \partial_t P(x_k,-)&=&d^-(x_{k}+a) P(x_k+a,-) -d^-(x_{k}) P(x_k,-) \nonumber\\
\fl &-&\frac{\alpha^-(x_{k})}2 P(x_k,-) + \frac{\alpha^+(x_{k})}2 P(x_k,+).
\end{eqnarray}
We are interested in cases where the typical run length is much longer
that the lattice spacing so that $d^\pm\gg\alpha^\pm$. Furthermore, when $a\rightarrow0$ while $v^\pm(x)=d^\pm(x) a$ remains finite, one gets, at leading order,
\begin{eqnarray}
\fl    \partial_t P(x,+)&=& - \grad[v^+(x) P(x,+)] +\frac{\alpha^-(x)}2 P(x,-) - \frac{\alpha^+(x)}2 P(x,+)+{\cal O}(a)\nonumber\\
\fl    \partial_t P(x,-)&=& \grad [v^-(x) P(x,-)] -\frac{\alpha^-(x)}2 P(x,-) + \frac{\alpha^+(x)}2 P(x,+)+{\cal O}(a),\nonumber\\\fl
\end{eqnarray}
which is exactly the master equation for run-and-tumble bacteria considered previously off lattice~\cite{Schnitzer1993,Tailleur2008}. Following the same path as there would lead to a Langevin
equation for the density of a large but finite number of bacteria
\begin{equation}
  \label{eqn:langdensNI}
  \dot \rho(x,t)=-\grad[\rho V -D \grad \rho + \sqrt{2 D \rho}\, \eta],
\end{equation}
where
\begin{eqnarray}
  \label{eqn:coeff}
   &&D=\frac{{\cal D}}{1+\xi_1};\qquad  V= \frac{\cal V}{1+\xi_1};\qquad  {\cal V}= \frac{\alpha^- v^+-\alpha^+ v^-}{2\alpha} - \frac v \alpha \grad\frac{v^+ v^-}{v} \nonumber\\
   &&  {\cal D}= \frac{v^+ v^-}{\alpha} ;\qquad\xi_1 = \frac{v^+}{2\alpha} \grad \frac{v^+}{v} -\frac{v^-}{2 \alpha} \grad\frac{v^-}v,
\end{eqnarray}
with $\alpha=(\alpha^++\alpha^-)/2$, $v=(v^++v^-)/2$. It was shown that~\eqref{eqn:langdensNI} captures the steady state of the off lattice model exactly~\cite{Tailleur2008} and in section~\ref{sec:FHIB} we will show that it also describes the large scale behaviour of run-and-tumble bacteria on lattice. The condition for
\eqref{eqn:langdensNI} to derive from an effective free energy is that there
exists an excess free energy functional ${\cal F}_{ex}[\rho]$ such that
\begin{equation}
  \frac{V}{D}=-\grad \frac{\delta {\cal F}_{ex}}{\delta \rho(x)},
\label{eq:FreeEnergyCond}
\end{equation}
which can be solved to give
\begin{equation}
  {\cal F}_{ex}[\rho]= \int_0^L \dx \left\{\rho(x)\left[\log\left(\frac{v^+ v^-}v\right)+\frac 1 2 \int_0^x\dx' \left(\frac {\alpha^+}{v^+}-\frac{\alpha^-}{v^-} \right)\right]\right\},
\label{eq:ExcessFreeEnergy}
\end{equation}
 as long as, for periodic boundary conditions, $\int_0^L \dx (\alpha^- v^+-\alpha^+ v^-)=0 $. The total free energy is then given by
\begin{equation}
  {\cal F}[\rho]= \int \dx \rho(\log\rho-1)+{\cal F}_{ex}[\rho].
\label{eq:TotalFreeEnergy}
\end{equation}
Note that it is not always possible to write a free energy of this form for non-interacting particles in higher dimensions, nor, in general, for interacting systems. As we can see, there is no gradient term in this expression. This is
due to the fact that when deriving the continuum limit, terms of order $a$ and higher are neglected. If equation~\eqref{eqn:langdensNI} leads to large gradients in the density, these higher order terms should be included and may alter the result; these terms control the surface tension, for example. These higher order terms could also violate the condition~\eqref{eq:FreeEnergyCond}.

We will now illustrate the steady state predictions of equations~\eqref{eq:LatticeNIProbA} \eqref{eq:LatticeNIProbB} and~\eqref{eq:TotalFreeEnergy} for a few simple cases and show the difference between the results predicted by the lattice model and by the continuum approximation.

\subsubsection{Examples}
Where not stated otherwise the simulations we present here use reflecting boundary conditions; if a particle tries to jump off one end of the lattice it is, instead, kept where it is but turned around. All simulations are performed with continuous-time Monte Carlo algorithms.
\subsubsection*{Position Dependent Rates with Closed Boundaries}
\begin{figure}
 \centering
 \begin{center}
  \includegraphics[width=7cm]{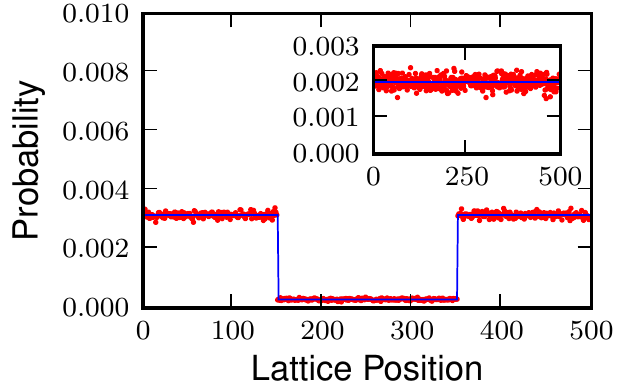}
 \end{center}
\caption{{\bf Main:} Steady state probability distribution for constant tumble rates, $\alpha^\pm_i=1$ and isotropic jump rates $d^\pm_i=1+10\,\theta(i-150)\,\theta(350-i)$. Data averaged from the positions of $400,000$ particles. {\bf Inset:} Steady state probability distribution for constant jump rates, $d_i^\pm=10$, and tumble rates $\alpha^\pm_i=\theta(i-150)\,\theta(350-i)$. Data from $100,000$ particles. In both figures simulation data are shown in red and the theory prediction in blue. Both simulations performed on a lattice of $500$ sites and recorded at $t=5000$.}
\label{fig:PositionDependent}
\end{figure}

First, we consider the case of a position dependent, but isotropic, jump rate and a constant tumbling rate. As a simple example we use a top-hat function for jump rate such that $d^\pm_i= 1+10\,\theta(i-150)\,\theta(350-i)$, where $\theta(x)$ is the Heaviside step function. Both the continuum and lattice theory predict that the average occupancy should be inversely proportional to the velocity,
\begin{equation}
 \rho(x)\propto \frac{1}{v(x)}; \qquad \rho_i=\rho_i^++\rho_i^-\propto \frac{1}{d_i}.
\end{equation}
 The results of the simulations and both predictions are shown in figure~\ref{fig:PositionDependent} (main).

In contrast to the jump rate, simply making the tumble rate depend on position but maintaining isotropy has no effect on the predicted distribution. Note that the free energy in equations~\eqref{eq:ExcessFreeEnergy} and \eqref{eq:TotalFreeEnergy} has no dependency on $\alpha$ for isotopic rates. Using the same form as for the position dependent jump rate in our simulations this can be verified numerically, see figure~\ref{fig:PositionDependent} (inset).

\subsubsection*{Direction Dependent Rates with Closed Boundaries}
In many physical situations, however, bacteria do not move unbiasedly but are affected by their external conditions. This may be due, for example, to sedimentation due to gravity, where there is an asymmetry in jump rates between left and right (or up and down) moving particles. Another case of interest may be anisotropic tumble rates. Bacteria undergoing chemotaxis often vary their tumble rate dependent on whether they are travelling up or down a chemical gradient. Though a simple asymmetry in tumble rate does not fully capture this behaviour, we do see particles preferentially move in the direction of a lower tumble rate, as would be expected.

These two cases show qualitatively the same behaviour, with an exponential decay in the unfavoured direction. From equations~\eqref{eq:LatticeNIProbA} and \eqref{eq:LatticeNIProbB}, the probabilities for left or right going particles on lattice are
\begin{equation}
 P(i,\pm) \propto \exp\left[i\,\log\left(\frac{d^+\left(2\,d^-+\alpha^-\right)}{d^-\left(2\,d^++\alpha^+\right)}\right)\right]\equiv \exp\left[i\,\lambda_{\rm latt}\right],
\end{equation}
while in the continuum case the probability density is given by
\begin{equation}
\rho(x)\propto\exp\left[x\frac{\alpha^-\,v^+-\alpha^+\,v^-}{2\,v^+\,v^-}\right]\equiv\exp\left[x\,\lambda_{\rm cont}\right].
\end{equation}
For expositional simplicity we consider homogeneous rates here. To examine the difference between these two cases, consider, for example, the case of sedimentation, where $d^\pm=d_0(1\pm\epsilon)$ and $\alpha^\pm=\alpha_0$. The decay constant in the continuum limit is then
\begin{equation}
 \lambda_{\rm cont}=\frac{\alpha_0\,\epsilon}{v_0(1-\epsilon^2)},
\label{eq:ContDecayConst}
\end{equation}
and the lattice decay constant is given by
\begin{equation}
 \lambda_{\rm latt}=\frac{\alpha_0\,\epsilon}{d_0(1-\epsilon^2)}-\frac{\epsilon\,\alpha_0^2}{2\,d_0^2(1-\epsilon^2)^2}+\mathcal{O}\left(\frac{\alpha_0}{d_0}\right)^3.
\label{eq:LattDecayConst}
\end{equation}
We can see then that the two decay lengths will be equal if the jump rate is much larger than the tumble rate, i.e. for large average run lengths. Both decay constants tend to zero as the asymmetry disappears, $\epsilon\rightarrow0$, but the ratio $\lambda_{\rm latt}/\lambda_{\rm cont}$ remains finite.

\begin{figure}
 \centering
 \begin{center}
  \includegraphics[width=7cm]{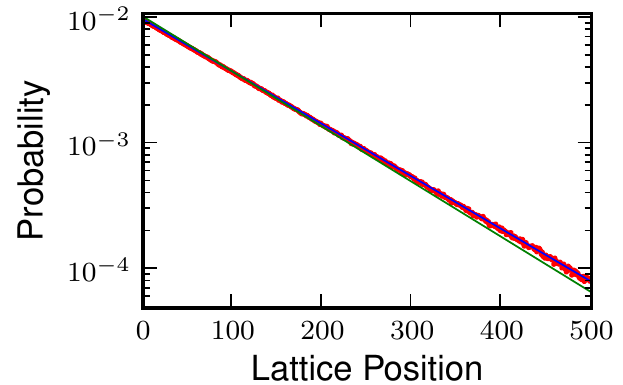}
 \end{center}
\caption{Steady state probability distribution for constant tumble rates, $\alpha^\pm_i=1$ and jump rates $d^\pm_i=10\mp1$. Simulation data are shown in red, the lattice prediction in blue and the continuum prediction in pink. Data collected from $10^7$ particles at $t=2000$ on a lattice of $500$ sites.}
\label{fig:DirectionDependent}
\end{figure}

In our simulations we use $d^\pm_i=10\mp1$ and $\alpha_0=1$, so the drift velocity, the external bias, is much less than the self-propelled speed, i.e. $\abs{d_i^+-d_i^-}\ll d^\pm$. The continuum theory predicts the distribution to be $\rho(x)\propto \exp(-x/99)$, while the lattice theory predicts $P(i)\propto (207/209)^i$. Both predictions are shown, along with the simulation data in figure~\ref{fig:DirectionDependent}. The ratio of decay constants is 
\begin{equation} 
\frac{a\,\lambda_{\rm latt}}{\lambda_{\rm cont}}=1-\frac{\alpha_0}{d_0(1-\epsilon^2)}+\mathcal{O}\left(\frac{\alpha_0}{d_0}\right)^2 \approx 0.95. 
\end{equation}
Note that in equations~\eqref{eq:ContDecayConst} and~\eqref{eq:LattDecayConst} one constant multiplies lattice position, $i$, and the other the continuum position, $x$, hence the factor of $a$. We see that the difference between the lattice and continuum results vanishes in the infinite run length limit, $d_0/\alpha_0\rightarrow\infty$, unless $\epsilon=1$, in which case both decay constants diverge.

\subsubsection*{Periodic Boundary Conditions}
We can also calculate the expected probability distribution for periodic boundary conditions. In this case our calculation on lattice is slightly more complicated as we do not know the current a priori, but must determine it through the conditions $P(L+1,\pm)=P(1,\pm)$ and $\sum_i\big[P(i,+)+P(i,-)\big]=1$.

We can write $P(i,+)=C_1(i)\big[P(1,+)-J\,C_2(i)\big]$, where $C_1(i)$ and $C_2(i)$ can be read from equation~\eqref{eq:LatticeNIProbA}. Then from the periodicity of the system we can write the current as
\begin{equation}
 J=\frac{C_1(L+1)-1}{C_1(L+1)\,C_2(L+1)}P(1,+)\equiv C_3 P(1,+).
\end{equation}
We can then use the normalisation of the distribution to determine $P(1,+)$ as
\begin{equation}
 P(1,+)=\frac{1}{\sum_{i=1}^L\left[\frac{d_i^++d_i^-+\alpha_i^++\alpha_i^-}{d_i^++d_i^-}C_1(i)\left(1-C_3\,C_2(i)\right)+\frac{C_3}{d_i^-+\alpha_i^-}\right]}.
\end{equation}

\begin{figure}
 \centering
\includegraphics[width=7cm]{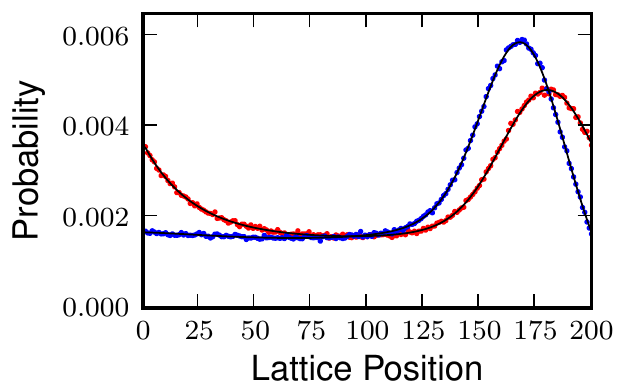}
\caption{The probability distributions at steady state for rates given by equations~\eqref{eq:PositionDependentRatesPBC}. The distribution for left moving particles is shown in blue and for right moving particles in red. The points show data from stochastic simulations and the solid lines the theoretical prediction. Data from $2,000,000$ particles at $t=200$ on a lattice of $200$ sites.}
\label{fig:PositionDependentRatesPBC}
\end{figure}

As an example see figure~\ref{fig:PositionDependentRatesPBC} where we consider the case where the jump and tumble rates are
\begin{equation}
 \begin{array}{rclrcl}
d_i^+ &=&  10 & d^-_i &=& \frac{2}{\exp(-(x-100)^2/5000)} \\
\alpha_i^+ &=&  1 &  \alpha_i^-&=&1.
\end{array}
\label{eq:PositionDependentRatesPBC}
\end{equation}
We omit the exact forms\ for the probability distributions as these do not reduce to a compact form and are not enlightening in themselves.

\subsection{A Zero-Range Interaction}
\label{sec:ZeroRange}
Though there is no generic solution for the steady state of interacting run-and-tumble particles, there are limiting cases that can be solved exactly, most simply a zero range interaction. We define the number of left and right going particles on a site as, respectively, $n^-_i$ and $n_i^+$. The total occupancy of a site is then $n_i=n_i^++n_i^-$. The simplest interaction we can add is a zero-range interaction, where the rate for a particle to jump from site $i$ with occupancy $(n_i^+,n_i^-)$ to site $i\pm1$ is defined as $d_i^\pm(n_i^+,n_i^-)$ and is a function of the number of particles at the departure site but not dependent on the number of particles at the arrival site. With this addition we can now see more complex behaviour and non-trivial steady states, even for homogeneous and isotropic jump and tumble rates, but can still, under certain conditions presented below, calculate the stationary probability distribution exactly.

As is standard for zero-range processes (ZRPs)~\cite{Evans2005}, we begin by assuming there exists a factorised form for the steady state probability distribution of the form
\begin{equation}
 P\left(\left\lbrace n_i^+,n_i^-\right\rbrace\right) \propto \prod_{j=1}^L g_j(n_j^+,n_j^-).
\end{equation}
This ansatz can then be substituted into the master equation for $P\left(\left\lbrace n_i^+,n_i^-\right\rbrace\right)$,
 \begin{eqnarray}
\label{eqn:mastereqZRP}
\fl    \partial_t P&=&\sum_{k=1}^L (n_k^++1)\,d_k^+(n_k^++1,n_k^-)\,P(n_k^++1,n_{k+1}^+-1)
    -n_k^+\,d_k^+(n_k^+,n_k^-)\,P(n_k^+,n_{k+1}^+)\nonumber
    \\\fl&+&(n_{k+1}^-+1)\,d_{k+1}^-(n_{k+1}^+,n_{k+1}^-+1)\,P(n_k^--1,n_{k+1}^-+1)
    -\frac{\alpha_k^-}{2} n_k^-P(n_k^-,n_k^+)\nonumber\\\fl
    &-&n_{k+1}^-\,d_{k+1}^-(n_{k+1}^+,n_{k+1}^-)\,  P(n_k^-,n_{k+1}^-)+\frac{\alpha_k^-}{2}(n_k^-+1)P(n_k^-+1,n_k^+-1)\nonumber\\\fl
    &+&\frac{\alpha_k^+}{2}(n_k^++1)P(n_k^--1,n_k^++1)-\frac{\alpha_k^+}{2}n_k^+P(n_k^-,n_k^+).
\end{eqnarray}
 Then, for periodic boundary conditions, one way we may choose to solve this equation is to separately balance the fluxes for right moving particles entering and exiting each site, left moving particles entering and exiting each site and particles tumbling between species on each site. We then arrive at three sufficient conditions on the allowed rates for such a factorised form to exist:
\begin{eqnarray}
 g_i(n^+,n^--1) &=& c\, n^-\,d_i^-(n^+,n^-)\,g_i(n^+,n^-) \label{eq:ZRPconditionsA}\\
  g_i(n^+-1,n^-) &=& c^\prime\, n^+\,d_i^+(n^+,n^-)\,g_i(n^+,n^-) \label{eq:ZRPconditionsB}\\
\fl  \;\,\quad g_i(n^+,n^-)\,n^-\,\alpha_i^-(n^+,n^-) &=& g_i(n^++1,n^--1)\,(n^++1)\,\alpha_i^+(n^++1,n^--1),
\label{eq:ZRPconditionsC}
\end{eqnarray}
in which $c$ and $c^\prime$ are arbitrary constants. The first two of these conditions are the same as Evans and Hanney found for their two species model without transmutation~\cite{Evans2003,Evans2005}, while the third is due to balancing the tumbling. Note that, in principle, there may be other ways in which we can balance these terms which could arrive at different conditions on the rates.

Putting the three conditions~\eqref{eq:ZRPconditionsA}-\eqref{eq:ZRPconditionsC} together and eliminating the factors $g_i(n^+,n^-)$ yields two constraints on our choice of rates:
\begin{eqnarray}
  d^-_i\left(n^+,n^-\right)\,d^+_i\left(n^+,n^--1\right) & =& d^-_i\left(n^+-1,n^-\right)\,d^+_i\left(n^+,n^-\right)\\
\frac{n^+\,\alpha_i^+\left(n^+,n^-\right)}{(n^-+1)\alpha_i^-\left(n^+-1,n^-+1\right)} & =& \frac{c}{c^\prime}\frac{(n^-+1)d^-_i\left(n^+-1,n^-+1\right)}{n^+\,d^+_i\left(n^+,n^-\right)}.
\end{eqnarray}

One natural, but again not necessary, way to fulfil these conditions is to take
\begin{eqnarray}
d_i^+(n^+,n^-) &=&u^+_i(n^+)\,\omega_i(n^++n^-) \nonumber\\
 d^-_i(n^+,n^-) &=&u^-_i(n^-)\,\omega_i(n^++n^-) 
 \label{eq:JumpConditions}
\end{eqnarray}
for the jump rates. That is, for the rate at which a left or right oriented particle moves to be a product of a function of the number of particles oriented in that direction, and a function of the total number of particles on a site. Both functions can vary from site to site; the first can also depend on the particle species, but the second must be the same for both. 

A sufficient conditions on the tumble rates is then to take
\begin{eqnarray}
  \alpha_i^+(n^+,n^-) & =& c u^+_i(n^+)\,A_i(n^++n^-) \\
 \alpha^-_i(n^+,n^-) & =&c^\prime u^-_i(n^-)\,A_i(n^++n^-). 
  \label{eq:TumbleConditions}
\end{eqnarray}
The functions $u^+_i(n^+)$ and $u^-_i(n^-)$ are the same as in~\eqref{eq:JumpConditions}; $A_i(r)$ is a new, unconstrained, function that appears in both rates.

Up to a constant that can be subsumed into the normalisation, the marginals are given as
\begin{equation}
\fl g_i(n^+,n^-) = \gamma^{n^-}\prod_{j=1}^{n^+}\! \frac{1}{jd^+_i(j,n^-)} \prod_{k=1}^{n^-}\! \frac{1}{kd^-_i(0,k)} =  \gamma^{n^-}\prod_{r=1}^{n}\!\frac{1}{\omega_i(r)}\prod_{j=1}^{n^+}\! \frac{1}{ju^+_i(j)} \prod_{k=1}^{n^-}\! \frac{1}{ku^-_i(k)},
\end{equation}
where $\gamma=c/c^\prime$ and $n=n^++n^-$. We can then re-write the probability of a given configuration as
\begin{equation}
 P\left(\left\lbrace n_i^+,n_i^-\right\rbrace\right) = \frac{1}{Z} \prod_{i=1}^{L} g_i(n_i^+,n_i^-) = \frac{1}{Z} e^{\sum_{i=1}^{L} \ln\left(g_i(n_i^+,n_i^-)\right)},
\end{equation}
 where $Z$ is a normalisation, and define an effective single site free energy $f_i(n_i^+,n_i^-)=-\log(g_i(n_i^+,n_i^-))$. Note that this is independent of $\alpha^\pm_i(n^+,n^-)$; the way in which we chose to solve the master equation does not lead to factorised steady states for which the distribution can depend on the tumbling rates. As we saw that asymmetric tumble rates could affect the equilibrium distribution for the non-interacting case, we might suppose there are other solutions for the zero-range process which admit distributions dependent on the tumble rates. Whether or not these allow for factorised steady states remains to be determined.

\begin{figure}
 \centering
\begin{minipage}{0.48\textwidth}
\centering
\includegraphics[width=7cm]{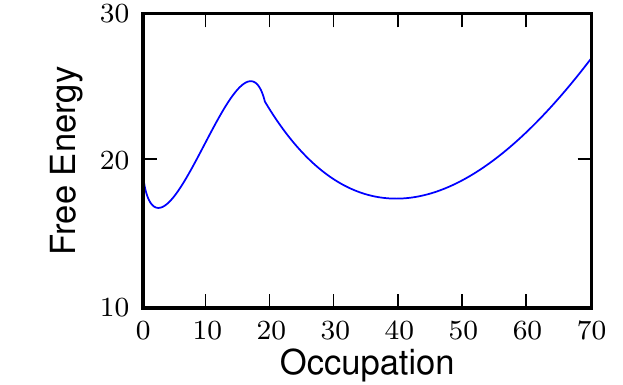}
\end{minipage}
\begin{minipage}{0.48\textwidth}
\centering
\includegraphics[width=7cm]{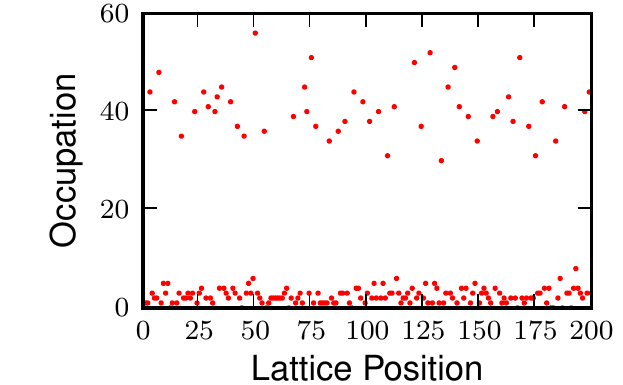}
\end{minipage}
\caption{{\bf Left}: The effective free energy density for the zero-range interaction with jump rate given by equation~\eqref{eqn:ZRPJumpRate} for $\alpha^\pm_i=1$, $\avg{n_i}=12$, $n_{m}=20$ and $v_0=10$. {\bf Right}: A typical snapshot of the system during its relaxation towards equilibrium for the same parameters on a lattice of $200$ sites and $2400$ particles at $t=1000$.}
\label{fig:ZRPSteadyState}
\end{figure}

To foreshadow the finite range interaction we will examine in section~\ref{sec:FHIB}, and to mimic the situation where an increase in density decreases the particles motility (as, for example, they get in each other's way) we now consider the following particular form of the steady state for this two-species ZRP for jump rates
\begin{equation} 
\label{eqn:ZRPJumpRate}
d^\pm_i(n^+,n^-)=\left\lbrace\begin{array}{ll} v_0\,\big[1-(n^++n^-)/n_{m}\big] &\qquad \mbox{if  } n^++n^-<n_{m} \\
                             v_0/n_{m} &\qquad \mbox{if  } n^++n^-\geq n_{m}
                            \end{array}\right.
\end{equation}
and tumble rate $\alpha^\pm_i=\alpha$. That is, the tumble rate is constant per particle and the jump rate decreases linearly as density increases until reaching a constant rate of $v_0/n_{m}$ at $n^++n^-=n_{m}-1$. In this case the effective free energy is double welled and the system separates into isolated sites of high and low density. The relative numbers of high and low density sites to which the system first separates are initial condition dependent. The system then relaxes via a series of evaporations and condensations towards a fixed steady state. This is reminiscent of what happens for single-species zero-range processes with similar jump rates. Where the jump rates remain finite even for very large occupancies the normal condensation is arrested~\cite{Thompson2010}. The free energy is shown in figure~\ref{fig:ZRPSteadyState} along with a typical snapshot of the system.

\begin{figure}
 \centering
 \includegraphics[width=7cm]{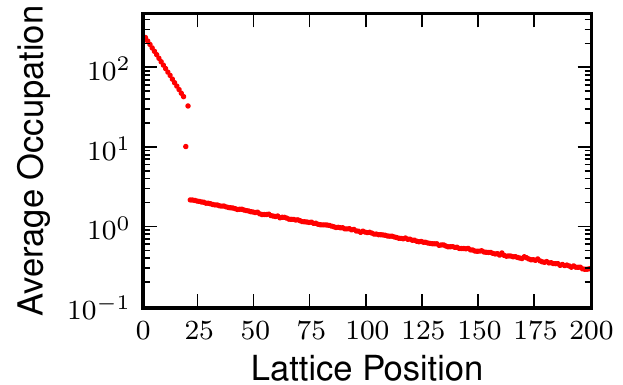}
 \caption{A time averaged snapshot of the steady state of zero-range process with parameters $n_{m}=20$, $\avg{n}=12$ $d^+_i=9$, $d^-_i=11$, $\alpha_i^\pm=1$. Data averaged from $10,000$ snapshots between $t=40,000$ and $t=50,000$}
 \label{fig:ZRPSedimentation}
\end{figure}

To this behaviour we can then add a drift term to simulate sedimentation by biasing the jump rates in one direction and applying closed boundary conditions. We see all the high density sites collect at one end of the lattice and the low density sites at the other, see figure~\ref{fig:ZRPSedimentation}. 

\section{Fluctuating Hydrodynamics for Interacting Bacteria}
\label{sec:FHIB}
We now turn to the more general case of $N$ interacting bacteria with interactions that are not limited to being on-site.
Specifically, we allow the jump and tumble rates to depend on the occupation numbers of each lattice site so that
\begin{equation}
 d_i^\pm = d_i^\pm(\bar{n}_i^\pm); \qquad \alpha^\pm_i = \alpha^\pm_i(\bar{n}^\pm_i),
\end{equation}
where $\bar{n}^\pm_i$ is a coarse-grained occupancy that depends linearly but non-locally on the occupancies of the whole lattice, 
\begin{equation}
 \bar{n}^\pm_i = \sum_j K^\pm_{i-j} n_j.
 \label{eqn:nmDefinition}
\end{equation}
In general the coarse graining kernel $K_{i-j}^\pm$ could also be a function of lattice position though here we do not consider that situation, where the manner in which the density is felt by the particles varies with position.
%The master equation for the probability of a configuration, $P(\left\lbrace n_k^+,n_k^-\right\rbrace)$, then reads
%\begin{equation}
%\label{eqn:mastereqFinite}
%  
%    \partial_t P&=\sum_{k=1}^L d_{k}^+ (n_k^++1) (1-\frac{n_{k+1}-1}{n_{m}})P(n_k^++1,n_{k+1}^+-1)
%    -d_k^+ n_k^+ (1-\frac{n_{k+1}}{n_{m}}) P(n_k^+,n_{k+1}^+)
%    \\&+d_{k+1}^- (n_{k+1}^-+1) (1-\frac{n_{k}-1}{n_{m}})P(n_k^--1,n_{k+1}^-+1)
%    - d_{k+1}^- n_{k+1}^- (1-\frac{n_{k}}{n_{m}})  P(n_k^-,n_{k+1}^-)\\
%    &+\frac{\alpha_k^-}{2}(n_k^-+1)P(n_k^-+1,n_k^+-1)-\frac{\alpha_k^-}{2} n_k^-P(n_k^-,n_k^+)\\
%    &+\frac{\alpha_k^+}{2}(n_k^++1)P(n_k^--1,n_k^++1)-\frac{\alpha_k^+}{2}n_k^+P(n_k^-,n_k^+)
%  
%\end{equation}
We aim at describing large scale behaviour, i.e. on a colony size, and so to derive a fluctuating hydrodynamics. In addition, this will allow us to compare again the phenomenology on and off lattice and to look for cases in which there is a ``free-energy'' like description, and for which we can thus characterise the steady state.
%Unlike in the zero-range case, we cannot now make a simple ansatz to solve the master equation exactly but we can still look for a free-energy like description of the steady state distribution. 
We follow a field theoretic approach to derive a continuum Langevin equation for the system, from which we can deduce the appropriate Fokker-Planck equation and the steady-state distribution.

\subsection{Field Theory for Non-interacting Particles}
\label{sec:NIFH}
To illustrate the technique we shall use to construct the fluctuating hydrodynamics for the full interacting case, let us first handle the non-interacting case and re-derive equation~\eqref{eqn:langdensNI}.

Field theoretic representations of lattice gases using bosonic coherent states were established in the 1970s following Doi and Peliti~\cite{Doi1976,Peliti1985}. The case where each site is limited to a single particle can be handled in some cases in this formalism~\cite{Wijland2001}, while more general finite occupancies could be handled using spin coherent states~\cite{Tailleur2008a}. Alternatively, probabilistic approaches from mathematical physics have also been used~\cite{Bertini2001,Pilgram2003,Jordan2004}. Here we use an alternative derivation, based on an approach \`{a} la Jansen and De~Dominicis~\cite{Janssen1976,DeDominicis1976} transposed in the context of the master equation. This is very similar to the generating function approach used by Biroli and Lefevre~\cite{Lefevre2007}. Beginning with a process discrete in both time and space one writes the probability of a trajectory as
\begin{eqnarray}
\fl P[\left\lbrace n_i^+(t_j),n_i^-(t_j)\right\rbrace] &=&\bigg<\prod_{i=1}^L\prod_{j=1}^{N}\delta\big(n^+_i(t_{j+1})-n^+_i(t_j)-J^+_i(t_j)\big)\nonumber\\
\fl&\times&\delta\big(n^-_i(t_{j+1})-n^-_i(t_j)-J^-_i(t_j)\big)\bigg>_{\!\vec{J}},
\end{eqnarray}
where $n^\pm_i(t_j)$ is the number of right ($+$) of left ($-$) moving particles at site $i$ at time $t_j$, the $J_i(t_j)^\pm$ are the changes in the number of each type of particle at each site at each time step and the bold faced $\vec{J}$ denotes the average is over all $J$'s. Re-writing the Dirac delta functions using imaginary Fourier representations this can be written as
\begin{eqnarray}
 \fl P[\left\lbrace n_i^+(t_j),n_i^-(t_j)\right\rbrace] &=&\!\int \prod_{i=1}^L\prod_{j=1}^{N}\! {\rm d}\hat{n}_i^+(t_j) {\rm d}\hat{n}_i^-(t_j) \bigg<\!\!\exp\!\Big(\hat{n}_i^+(t_j)\big(n^+_i(t_{j+1})-n^+_i(t_j)-J^+_i(t_j)\big)\nonumber\\
\fl &+&\hat{n}_i^-(t_j)\big(n^-_i(t_{j+1})-n^-_i(t_j)-J^-_i(t_j)\big)\Big)\bigg>_{\vec{J}},
\end{eqnarray}
where it should be noted that the conjugate fields $\hat n_i^\pm$ are imaginary. The average over the $J$'s can then be calculated explicitly from the dynamics. Specifically, a right moving particle can jump from site $i$ to site $i+1$ at time step $j$ with probability $n^+_{i}(t_j)d_i^+\dt$, where $\dt$ is the duration of a time step. The corresponding values of the $J$'s are $J_i(t_j)=-1$ and $J_{i+1}(t_j)=1$. Calculating all other moves and the probability that nothing happens, which corresponds to all $J_i^\pm=0$, we can write
\begin{eqnarray}
 \avg{e^{-\hat{n}_i^+(t_j)J^+_i(t_j)-\hat{n}_i^-(t_j)J^-_i(t_j)}}_{\vec{J}} &=&  1+n_i(t_j)^+d_i^+\left(e^{\hat{n}_i^+(t_j)-\hat{n}_{i+1}^+(t_j)}-1\right)\dt\nonumber\\
&+&n_{i+1}^-(t_j)d_{i+1}^-\left(e^{\hat{n}^-_{i+1}(t_j)-\hat{n}_i^-(t_j)}-1\right)\dt\nonumber\\ &+&\frac{\alpha_i^+}{2}n^+_i(t_j)\left(e^{\hat{n}^+_i(t_j)-\hat{n}^-_i(t_j)}-1\right)\dt\nonumber\\
&+&\frac{\alpha_i^-}{2}n_i^-(t_j)\left(e^{\hat{n}_i^-(t_j)-\hat{n}_i^+(t_j)}-1\right)\dt.
\end{eqnarray}
As this is of the form $1+k\dt$ we can approximate it as $\exp(k\dt)$ and write the probability for the trajectory as
\begin{eqnarray}
\fl  P[\left\lbrace n_i^+(t_j),n_i^-(t_j)\right\rbrace] &= & \int \Bigg(\prod_{i=1}^L\prod_{j=1}^{N} {\rm d}\hat{n}_i^+(t_j) {\rm d}\hat{n}_i^-(t_j)\Bigg) \exp\Bigg[\sum_{i=1}^{L}\sum_{j=1}^{N}\nonumber\\ 
\fl&&\bigg(\hat{n}_i^+(t_j)\left(n_i^+(t_{j+1})-n_i^+(t_j)\right) +\hat{n}_i^-(t_j)\left(n_i^-(t_{j+1})-n_i^-(t_j)\right)\nonumber\\ \fl 
&& +n_i(t_j)^+d_i^+\left(e^{\hat{n}_i^+(t_j)-\hat{n}_{i+1}^+(t_j)}-1\right)\dt\nonumber\\
\fl&&+n_{i+1}^-(t_j)d_{i+1}^-\left(e^{\hat{n}^-_{i+1}(t_j)-\hat{n}_i^-(t_j)}-1\right)\dt\nonumber\\
\fl&&+\frac{\alpha_i^+}{2}n^+_i(t_j)\left(e^{\hat{n}^+_i(t_j)-\hat{n}^-_i(t_j)}-1\right)\dt\nonumber\\ \fl
&&+\frac{\alpha_i^-}{2}n_i^-(t_j)\left(e^{\hat{n}_i^-(t_j)-\hat{n}_i^+(t_j)}-1\right)\dt\bigg)\Bigg].
\end{eqnarray}
We can then take a continuous time limit and make the substitutions
\begin{equation}
\fl n_i^\pm(t_{j+1})-n_i^\pm(t_j)\rightarrow \dot{n}_i^\pm\dt;\qquad \sum_{j=1}^N \dt \rightarrow \int_0^{T=N\dt} \dt;\qquad \prod_{j=1}^N {\rm d}\hat{n}_i^\pm(t_j)\rightarrow \mathcal{D} \hat{n}_i^\pm.
\end{equation}
The probability of a trajectory can then be written
\begin{equation}
  \label{eqn:probaction}
  P[\left\lbrace n_i^+(t),n_i^+(t)\right\rbrace]=\int \prod_i{\cal D}[\hat n_i^+,\hat n_i^-] e^{-S[{\bf n^+},{\bf n^-},{\bf \hat n^+},{\bf \hat n^-}]},
\end{equation}
where the action $S$ is given by
\begin{eqnarray}
  \label{eqn:action}
\fl  S&=&-\int_0^T \dt \sum_i \bigg[ \hat n_i^+ \dot n_i^++\hat n_i^- \dot n_i^-+n_i^+
  d_i^+ \left(e^{\hat n_i^+-\hat n_{i+1}^+}-1\right) +n_{i+1}^-
  d_{i+1}^-\left(e^{\hat n_{i+1}^--\hat n_{i}^-}-1\right)\nonumber\\
\fl &+&\frac{\alpha_i^+}{2} n_i^+ \left(e^{\hat n_i^+-\hat n_{i}^-}-1\right)+\frac{\alpha_i^-}{2} n_i^-\left(e^{\hat n_i^--\hat n_{i}^+}-1\right)\bigg].
\end{eqnarray}
Note that generic changes of variables in \eqref{eqn:probaction} will result in Jacobians. If these do not depend on the fields, $n_i^\pm$ and $\hat n_i^\pm$, they can be subsumed into the normalisation of the path integral but they must be handled with care otherwise.

We further simplify by considering
symmetric, constant, rates $d_i^+=d_i^-=d$ and
$\alpha_i^+=\alpha_i^-=\alpha$; the more general case causes little conceptual difficulty but is considerably more cumbersome as an illustration. Let us then introduce the new variables
\begin{equation}
\fl  \rho_i=n_i^++n_i^-;\quad J_i=d(n_i^+-n_i^-);\quad \hat \rho_i=\frac 1 2 (\hat n_i^++\hat n_i^-);\quad \hat J_i=\frac 1 2 (\hat n_i^+-\hat n_i^-)
\end{equation}
The action can then be written as
\begin{eqnarray}
\fl    S&=&-\int_0^T \dt \sum_i \bigg[\hat \rho_i \dot \rho_i + \frac 1 d \hat J_i
    \dot J_i + \frac d 2 \rho_i \left(e^{-(\hat
    \rho_{i+1}-\hat \rho_i + \hat J_{i+1}-\hat J_i)}+
e^{\hat \rho_{i+1}-\hat \rho_i - (\hat J_{i+1}-\hat
      J_i)}-2\right)\nonumber\\
\fl    &+&\frac {J_i}2 \left(e^{-(\hat
    \rho_{i+1}-\hat \rho_i + \hat J_{i+1}-\hat J_i)} -
    e^{\hat \rho_{i+1}-\hat \rho_i - (\hat J_{i+1}-\hat
    J_i)}\right)\nonumber\\
\fl    &+&\frac {d}2(\rho_{i+1}-\rho_i-\frac{J_{i+1}-J_i}d)
    \left( e^{ \rho_{i+1}-\hat \rho_i - (\hat
    J_{i+1}-\hat J_i)}-1\right)\nonumber\\
\fl    &+&\frac {\alpha \rho_i}4 \left(e^{2 \hat J_i}+e^{-2 \hat J_i}-2\right)+\frac {\alpha J_i}{4d} \left(e^{2 \hat J_i}-e^{-2 \hat J_i}\right)\bigg].
\end{eqnarray}
The continuum limit can be taken by explicitly introducing the
lattice spacing $a$ and making the substitutions
\begin{eqnarray}
    &\rho_i\to a\rho(x);\quad \hat\rho_i\to\hat\rho(x);\quad d\to v a^{-1};\quad\sum_i
  \to \int_0^{\ell=L a} \dx a^{-1};\nonumber\\
  &\quad J_i\to J(x);\quad \hat J_i\to \hat J(x);\quad \grad_i \to a \grad +\frac 1 2  a^2 \Delta,
\end{eqnarray}
where $\grad_i$ is the discrete gradient, e.g. $\grad_i\rho_i=\rho_{i+1}-\rho_i$, and $\ell$ is the system length. After Taylor expanding the action in powers of the lattice spacing and taking a diffusive rescaling of time and space, see~\ref{app:Scaling}, we find that the fluctuating hydrodynamic action is given by
\begin{equation}
  S_0=-\int_0^\tau\dt\int_0^1 \dx\; \bigg(\hat \rho \dot \rho - v \rho \grad \hat J - J \grad \hat \rho +\alpha \rho \hat
  J^2+\frac{\alpha J \hat J}{v}\bigg),
\end{equation}
which is invariant under a further diffusive rescaling of space and time. 

Going back to the definition of the probability \eqref{eqn:probaction}, we can then work backwards to recover a continuity equation for $\rho$ from our action~\cite{Tailleur2008a}. Starting from
\begin{equation}
  P[\left\lbrace\rho(x,t),J(x,t)\right\rbrace]=\frac 1 {\tilde Z} \int {\cal D}[\hat \rho,\hat J] e^{-S_0[\rho,J,\hat \rho,\hat J]},
\end{equation}
one can remove the quadratic term $\hat J^2$ by introducing a new field $\eta(x,t)$ via a Hubbard-Stratonovich transformation so that
\begin{equation}
  P[\left\lbrace\rho(x,t),J(x,t)\right\rbrace]=\frac 1 {\tilde Z} \int {\cal D}[\hat \rho,\hat J,\eta] e^{-S_0[\rho,J,\hat \rho,\hat J,\eta]},
\end{equation}
where the new action now reads (after some integration by parts)
\begin{equation}
  S_0=-\int_0^\tau \dt \int_0^1 \dx\;\bigg( \hat \rho \dot \rho +  v\grad \rho \hat J +  \hat \rho \grad J +\sqrt{ 2 \alpha \rho}\, \eta \hat
  J+\frac{\alpha J \hat J}{ v} -\frac 1 2 \eta^2\bigg).
\end{equation}
The integral over $\hat \rho$ and $\hat J$ then leads to
\begin{equation}
  \label{eqn:resultFH}
\fl  P[\left\lbrace\rho(x,t),J(x,t)\right\rbrace]\propto   \int {\cal D}[\eta] \delta(\dot \rho + \grad J)\; \delta\Big(\frac \alpha v J + \sqrt{ 2 \alpha \rho} \,\eta + \grad v \rho \Big)\; e^{-\frac 1 2 \int \dx \dt \;\eta^2},
\end{equation}
where the delta functions impose the two dynamic field equations
\begin{equation}
    \dot \rho = -\grad J; \qquad J= - D \grad \rho + \sqrt{2 D \rho }\, \eta;\qquad D=\frac{v^2}{\alpha}.
\end{equation}
Given its weight in \eqref{eqn:resultFH}, $\eta(x,t)$ is a Gaussian white noise:
\begin{equation}
  \langle \eta(x,t) \eta (x',t')\rangle=\delta(x-x') \delta(t-t').
\end{equation}
This is consistent with the calculation off-lattice for non-interacting, homogeneous and isotropic systems and validates the results obtained previously, see equation~\eqref{eqn:langdensNI}.

\subsubsection{Fluctuating Hydrodynamics and Large Deviation Functions}
Before going any further, let us make a brief detour to consider the connection with the standard
fluctuating hydrodynamics approach considered in the mathematics literature~\cite{Bertini2004}. Let us first note that from the definition of the continuum limit, one has
\begin{equation}
N=\sum_i \rho_i=\int_0^1 \dx \rho(x)
\end{equation}
The integral of the density field is thus an extensive variable. On
the other hand, the density field considered by mathematicians is often defined by
\begin{equation}
  \rho(x)=\frac 1 \ell \sum_i \rho_i \delta(x-a i)
\end{equation}
and satisfies the normalisation condition
\begin{equation}
  \int \dx \rho(x) = \frac N \ell
\end{equation}
To make the connection between the two approaches, it is thus natural
to rescale our density term to make the extensivity apparent:$\rho \to
\ell \rho$. To ensure that the conservation equation still has the form
$\dot \rho = - \grad J$, one must also rescale the current field $J \to \ell
J$. Before introducing the $\eta(x,t)$ field, the action thus reads
\begin{equation}
  S_0=-\ell\int_0^\tau\dt \int_0^1 \dx\; \bigg(\hat \rho \dot \rho - v \rho \grad \hat J - J \grad \hat \rho +\alpha \rho \hat
  J^2+\frac{\alpha J \hat J}{ v}\bigg).
\end{equation}
One can again introduce the noise field and integrate over the conjugates fields $\hat \rho$ and $\hat J$ to get
\begin{equation}
\fl  P[\left\lbrace\rho(x,t),J(x,t)\right\rbrace]\propto   \int {\cal D}[\eta] \delta(\dot \rho + \grad J)\; \delta\Big(\frac \alpha v J + \sqrt{ 2 \alpha \rho} \,\eta + \grad v \rho \Big)\; e^{-\frac \ell 2 \int \dx \dt \;\eta^2}.
\end{equation}
Interestingly, the fields are now all intensive, and the smallness of
the noise does not come from a $\sqrt{\rho}$ versus $\rho$ noise prefactor,
but from its explicit variance, read in the Gaussian weight:
\begin{equation}
  \langle \eta(x,t) \eta (x',t')\rangle=\frac 1 \ell \delta(x-x') \delta(t-t').
\end{equation}
This is the usual fluctuating hydrodynamics, as considered, for instance,
in~\cite{Bertini2004}. In the large size
limit, the first order correction to the deterministic equation in a
$1/\ell$ expansion is given by the addition of the noise term $\sqrt{2 D
\rho}$. This noise is typically of order $1/\sqrt \ell$,
i.e. trajectories of probability of order 1 have $\ell \eta^2$ of order
1. Large deviations correspond to trajectories where the noise can be of
order one. They yield probabilities of order ${\cal
O}(\exp(-\ell))$, and are described by the fluctuating hydrodynamics constructed here.

\subsection{Interacting Particles}
\label{sec:stocheqI}
Consider now the case of interacting particles where the jump and tumble rates depend on the occupation numbers of each lattice site. Our velocity is then modified to
\begin{equation}
 v^\pm(x) \rightarrow v^\pm(\bar{\rho}^\pm(x),x),
\end{equation}
and the tumble rate to
\begin{equation}
  \alpha^\pm(x) \rightarrow \alpha^\pm(\bar{\rho}^\pm(x),x),
\end{equation}
with $\bar{\rho}^\pm(x)$ given by an integral over the density
\begin{equation}
 \bar{\rho}^\pm(x) = \int K^\pm(x-y) \rho(y) {\rm d}y.
\end{equation}

Following the same path as that followed in section \ref{sec:NIFH}
for the non-interacting case, one gets for the action
\begin{eqnarray}
\fl  S&= &-\int \dt \dx\;\bigg[ \hat \rho \dot \rho - \frac{v^+(\bar{\rho}^+)v^-(\bar{\rho}^-)}{v(\bar{\rho}^+,\bar{\rho}^-)} \rho \grad \hat J - J \grad \hat \rho +\frac{\alpha^+(\bar{\rho}^+)v^-(\bar{\rho}^-)+\alpha^-(\bar{\rho}^-)v^+(\bar{\rho}^+)}{v(\bar{\rho}^+,\bar{\rho}^-)} \rho \hat J^2\nonumber\\
\fl& +&\frac{\alpha(\bar{\rho}^+,\bar{\rho}^-)}{v(\bar{\rho}^+,\bar{\rho}^-)} J \hat J +\ell \frac{\alpha^+(\bar{\rho}^+)v^-(\bar{\rho}^-)-\alpha^-(\bar{\rho}^-)v^+(\bar{\rho}^+)}{v(\bar{\rho}^+,\bar{\rho}^-)}\rho \hat J\bigg],
\end{eqnarray}
where $v=(v^++v^-)/2$ and $\alpha=(\alpha^++\alpha^-)/2$. The factor of $\ell$ in the final term implies that, for the diffusive scaling to hold, at a scale $\ell$, the asymmetry $\alpha^+(\bar{\rho}^+)v^-(\bar{\rho}^-)-\alpha^-(\bar{\rho}^-)v^+(\bar{\rho}^+)$ must be of order $1/\ell$. This is reminiscent of the ASEP, where if the bias is much smaller than $1/\sqrt{\ell}$ the diffusive scaling holds (Edwards-Wilkinson universality class), as in the symmetric exclusion process, but for larger asymmetries the dynamic exponent $z$ is the same as Kardar-Parisi-Zhang scaling~\cite{Prolhac2009}. Integrating over $\hat \rho$ and $\hat J$ now yields the set of field equations
\begin{equation}
\label{eq:LEFieldTheory}
    \dot \rho = -\grad J; \qquad J= - D \grad \rho -V\rho + \sqrt{\frac{v(\alpha^+\,v^-+\alpha^-\,v^+)}{\alpha^2} \rho } \eta,
\end{equation}
with
\begin{equation}
  \label{eqn:coeffFT}
   D= \frac{v^+ v^-}{\alpha};\qquad  V= \ell \frac{\alpha^- v^+-\alpha^+ v^-}{2\alpha} - \frac v \alpha \grad\frac{v^+ v^-}{v}
\end{equation}

This formalism can now be used to analyse the effect of interactions on the large scale behaviour of a system of run-and-tumble particles.

\section{Crowding Interactions}
\label{sec:CrowdingInt}

Having set up and utilised a field theory apparatus to derive a fluctuating hydrodynamics for a general linear dependence on density in the jump rates, let us turn now to a specific class of interactions. In particular we consider a crowding interaction, where the velocity of the bacteria decreases with increasing density, in which case we expect to see our system separate into regions of high and low density as particles become trapped in regions of high density~\cite{Tailleur2008}. 

In general we expect to see qualitatively similar behaviour for any choice of $v(\rho)$ which decrease sufficiently quickly towards a finite non-zero velocity at high densities. In the following we use 
\begin{equation}
 v^\pm(\bar{\rho}^\pm)=\left\lbrace\begin{array}{ll} v_0\,\left(1-(\bar \rho^\pm)/\rho_{m}\right) &\qquad \mbox{if  } \bar \rho^\pm<\rho_{m} \\
                             v_0/\rho_{m} &\qquad \mbox{if  } \bar \rho^\pm \geq\rho_{m}
                            \end{array}\right.,
\label{eqn:CrowdingVelocity}
\end{equation}
as we did for the exactly solvable zero-range process.
  
\subsection{Zero-Range Interactions}
\label{sec:ContSpaceZRP}
This approach can describe many types of interaction, in particular let us now consider the zero range interaction we met in section~\ref{sec:ZeroRange} in the context of the fluctuating hydrodynamics we developed in section~\ref{sec:FHIB}. This provides us with a benchmark to check that our fluctuating hydrodynamics is consistent with the exact results we obtained previously.

For the zero-range process, where the velocity depends only on the occupation at the departure site, our kernel is given by $K_i^\pm=\delta_{i0}$ and in continuum $\bar \rho^\pm(x) = \rho(x)$. For simplicity, and to compare with our previous results we shall take $\alpha^\pm(\bar \rho)=\alpha$. This simplifies equation~\eqref{eq:LEFieldTheory} considerably and, indeed, guarantees that $\alpha^+v^--\alpha^-v^+=0$. From section \ref{sec:stocheqI} we know that the fluctuating hydrodynamics describing the run-and-tumble bacteria with velocity given in equation~\eqref{eqn:CrowdingVelocity} are given by
\begin{eqnarray}
   \dot \rho &=& -\grad J; \quad J= - D \grad \rho -\frac{v_0}{\alpha}
    \Big(1-\frac{\rho}{\rho_{m}}\Big)\grad\Big[ v_0\Big(1-\frac{\rho}{\rho_{m}}\Big)\Big]\rho + \sqrt{2
    D \rho } \eta;\nonumber \\ D&=&\frac{v_0^2\Big(1-\frac{\rho}{\rho_{m}}\Big)^2}{\alpha}.
\end{eqnarray}
The corresponding Fokker-Planck equation is given by
\begin{equation}
 \fl \dot{\cal P} = \int \dx \frac{\delta}{\delta\rho(x)} \partial_x\left[-\frac{v_0}{\alpha}
    \Big(1-\frac{\rho}{\rho_{m}}\Big)\partial_x\Big[ v_0\Big(1-\frac{\rho}{\rho_{m}}\Big)\Big]\rho -D\partial_x\rho - D\rho\left(\partial_x\frac{\delta}{\delta\rho(x)}\right)\right]{\cal P} \label{FPE}.
\end{equation}
Note that a term $\grad \frac{\delta}{\delta \rho}[D(\rho)]$ could be
present, but vanishes for symmetry reasons
(See~\cite{Tailleur2008,Garcia1999}). Looking for a free energy $P\propto
\exp[-{\cal F}[\rho]]$ one gets
\begin{equation}
  -\grad \frac{\delta {\cal F}}{\delta \rho}=-\grad\Big[\log \rho + \log \Big(\Big(1-\frac\rho{\rho_{m}}\Big)\Big)\Big],
\end{equation}
whose solution is
\begin{equation}
\fl  {\cal F}[\rho]=\int \dx \rho(\log\rho-1)-(\rho_{m}-\rho)\Big[\log\Big(1-\frac\rho{\rho_{m}}\Big)-1\Big],\qquad {\rm for }\, \rho<\rho_{m}.
  \label{eq:ZRPFreeEnergy}
\end{equation}
For $\rho\geq\rho_{m}$ the free energy density $f(\rho(x))$ is given by
\begin{equation}
f(\rho(x))= \rho\bigg(\log\left(\frac{\rho}{\rho_{m}}\right)-1\bigg),
\end{equation}
which corresponds precisely to the free energy calculated exactly in section~\ref{sec:ZeroRange} for the total occupancy and in the continuum limit. An example of this free energy for one choice of parameters is shown in figure~\ref{fig:ZRPSteadyState} (left).

\subsection{Finite Range Interactions}
\label{sec:FiniteRange}
We saw in section~\ref{sec:ContSpaceZRP} that our fluctuating hydrodynamics admit a free energy and accurately reproduced the exact results for the zero-range process. Let us now extend that analysis to a system with finite range interactions. As before we now take the coarse grained density $\bar \rho ^\pm$ to be given by
\begin{equation}
 \bar{\rho}^\pm(x) = \int K^\pm(x-y) \rho(y) {\rm d}y.
\end{equation}
For smooth profiles, we hope that the differences between $\bar \rho^\pm$ and $\rho$ are small so we can treat the free energy in equation~\eqref{eq:ZRPFreeEnergy} as a mean field theory. This way we can still use the free energy we derived in section~\ref{sec:ContSpaceZRP} to predict the coexistence densities and instability to spinodal decomposition. The finite range nature of the interactions will introduce correlations between sites which we hope will manifest only via surface tension~\cite{Tailleur2008}. We hope that this surface tension will only effect a clustering of the high and low density sites without affecting the coexistence densities. If the mean field theory captures the picture correctly, spinodal decomposition occurs whenever the second derivative of the free energy density is negative, i.e. for
\begin{equation}
  \frac 1 \rho + \grad \log v[\rho]<0 \Leftrightarrow \frac 1 \rho - \frac 1 {\rho_{m}-\rho}<0.
\end{equation}
Thus, whenever $\rho_{m}>\rho>\rho_{m}/2$, the system should be unstable with respect to spinodal decomposition.

\subsubsection{Isotropic Kernels}
\begin{figure}
 \centering
\includegraphics[width=7cm]{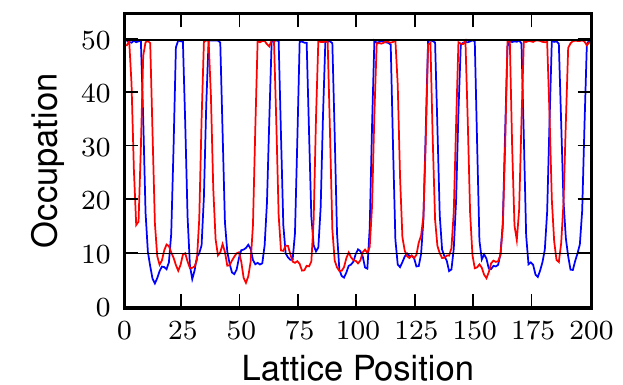}
\caption{Snapshot of typical density profiles for an average run length of $100$ sites (red) and $10$ sites (blue) for the isotropic, Gaussian kernel. The black lines show the predicted average high and low densities. Data recorded at $t=1000$ using $5000$ particles with $n_m=50$, $k=2$ and $\alpha=1$.}
\label{fig:SymmetricKernelSnapshot}
\end{figure}

If we use an isotropic kernel to calculate $\bar \rho^\pm$ in our jump rates we do indeed recover results consistent with the zero-range free energy. That is, if we take $K^+(x-y)=K^-(x-y)$ our simulations match the free energy predictions. In particular, we have worked with a Gaussian kernel, where $K^\pm(x)=\exp(-x^2/k)/Z$, with Z a normalisation and $k$ a parameter to control the range of the interaction. The results of simulations using this kernel are shown in figure~\ref{fig:SymmetricKernelSnapshot} along with the predicted average high and low densities from the free energy. To calculate the coexisting densities of the high and low density sites we form a double tangent construction on the free energy~\cite{Sollich2002}. As predicted the finite range nature of the interactions effectively creates a surface tension, but does not significantly alter the coexistence densities. Further, as expected we see no dependence in the steady state on our choice of $v_0$ and $\alpha$. 

\subsubsection{Anisotropic Kernels}
For anisotropic coarse graining kernels, however, the situation is more complex. One simple and natural way to introduce an anisotropic kernel is to account for the finite volume of bacteria by stating that there can be at most $n_{m}$ bacteria on each lattice site and taking the occupancy at the arrival site as our $\bar n$ in equation~\eqref{eqn:nmDefinition}. This forms a generalisation of the partial exclusion process~\cite{Tailleur2008a,Schutz1994} and results in the jump rates
\begin{equation}
 d_i^\pm(n^+_i,n^-_i) = d_i^\pm\left(1-\frac{n_{i\pm1}}{n_{m}}\right).
\label{eq:AnisotropicRate}
\end{equation}

\begin{figure}
\centering
  \includegraphics[width=7cm]{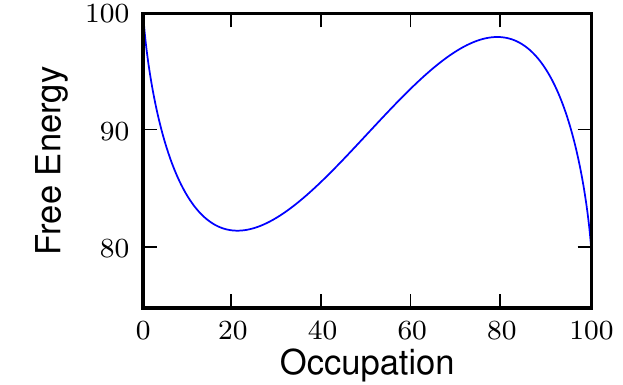}
  \caption{Free energy density of the exclusion model where the occupation of a site is constrained to be smaller or equal to $100$ particles by the choice of rates~\eqref{eq:AnisotropicRate}.}
\label{fig:ExclusionFreeEnergy}
\end{figure}

In this case the effective free energy is limited to a region where $\rho<\rho_{m}$ and this section is sketched in figure \ref{fig:ExclusionFreeEnergy}.
A double tangent construction amounts to looking for a density $\rho_{low}$
such that the tangent of the free energy density at this point
meets with the free energy density at $\rho=\rho_{m}$, as can be seen
from inspection of figure \ref{fig:ExclusionFreeEnergy}. This 
amounts  to finding $\rho$ such that
\begin{equation}
  2 \Big(\frac \rho{\rho_{m}}-1\Big) = \log\frac\rho{\rho_{m}},
\end{equation}
which can be solved numerically and yields for the low density $\rho_{low}$ that coexists with $\rho_m$
\begin{equation}
  \frac{\rho_{low}}{\rho_{m}}=.203188.
\end{equation}
For a total average density larger that $\rho_{low}$, the stable phase
should thus be a combination of two phases, one with $\rho=\rho_{low}$ and
one with $\rho=\rho_{m}$, the ratio of the amounts of the two phases being set by
constraint on the global mass.

\begin{figure}
 \centering
 \includegraphics[width=7cm]{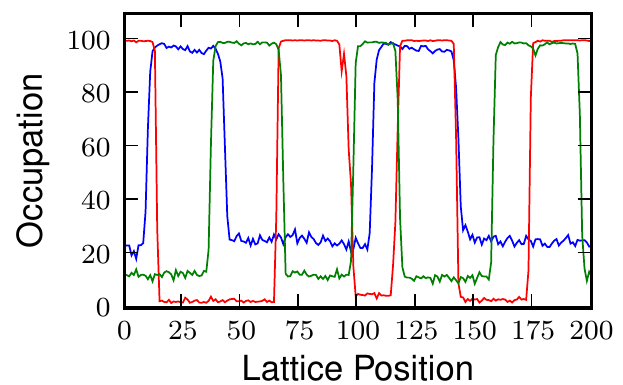}
 \caption{Snapshots of typical density profiles for an average run length of $100$ sites (red), $20$ sites (green) and $10$ sites (blue). Data recorded at $t=1000$ with $\alpha=1$, $n_m=100$ and from $10,000$ particles.}
 \label{fig:SimpleExclusionSnapshot}
\end{figure}

Interestingly, although the theory correctly predicts a change from a flat profile to phase separation, on examining the results of simulations of the underlying lattice system we found that the densities into which the system separates do not correspond to those predicted by the continuum theory. Indeed, while the continuum theory had no dependence on the tumble rate $\alpha$ or the coefficient of the jump rate $v_0$, the simulations for an anisotropic kernel showed a strong dependence on the ratio of these two parameters. The lower and upper densities both varied considerably with the average run length $r\equiv v_0/\alpha$, as shown in figure~\ref{fig:SimpleExclusionSnapshot}, and below $r=4$ we see no separation at all.

This discrepancy is not limited to the particular choice of anisotropic kernel we use as illustration above and is general to any anisotropic choice of $K^\pm(x)$. We have also conducted simulations with smooth anisotropic kernels without a hard limit on the number of particles per site, and saw exactly the same qualitative effect. Thus we have seen that the relevant factor is indeed the isotropy of the kernel but although the origin of the difference has been established, a comprehensive explanation for the variation between isotropic and anisotropic kernels is yet to be formulated.

Note that it is not in itself surprising that a fluctuating hydrodynamics developed to describe smooth profiles fails to quantitatively amount for the coexisting densities of profiles alternating high and low densities. One of the reasons our fluctuating hydrodynamics work so well for the isotropic case is that we always consider large occupancies on each lattice site. This means that the model is close to mean-field in the same sense as the large spin limit of a spin chain is well described by a continuous spin chain~\cite{Tailleur2008a}. Smaller occupancies would lead to quantitative differences between the predicted coexisting densities and those predicted from the fluctuating hydrodynamics, even for isotropic kernels. Furthermore, the Ito drift that was neglected in equation~\eqref{FPE} for symmetry reasons would not vanish for the anisotropic case. In fact even for the off-lattice model, the fluctuating hydrodynamics developed previously~\cite{Tailleur2008} is only valid for isotropic kernel and the quantitative mismatch between the fluctuating hydrodynamics and the simulations on lattice for anisotropic kernels are thus not that surprising. We nevertheless now try to shed some light on their origin.

%As in the zero-range case, we can also add a drift at this stage to mimic sedimentation, see fig~\ref{fig:ExclusionSedimentation}. The results bear a close similarity to those obtained from the purely local interaction, though the finite range of the exclusion does lead to correlations in the density absent in the zero-range case, see fig~\ref{fig:SedimentationCorrelations}.%

\subsubsection{Stability Analysis}
\label{sec:Stability}
One way we can analyse the difference between the isotropic and anisotropic interaction kernels is to examine the dynamic stability of the two systems. We consider a one dimensional system evolving from a homogeneous state under a small perturbation and determine whether the system is dynamically stable or unstable, whether the perturbations will, on average, grow or shrink. 

One possibility is that the discrepancies we saw in figure~\ref{fig:SimpleExclusionSnapshot} between the theory and simulations arise from the assumptions behind the diffusive limit taken in the field theory, see section~\ref{sec:FHIB} and~\ref{app:Scaling}. We therefore start from the continuum microscopic mean field equations for homogeneous and isotropic rates, i.e. after the continuum limit has been taken but before the diffusive limit,
\begin{eqnarray}
  \dot{\rho}^+ &=& -v\grad\left[\rho^+\left(1-\frac{\rho}{\rho_m}\right)\right]-\frac{\alpha\,\rho^+}{2}+\frac{\alpha\,\rho^-}{2} \\
  \dot{\rho}^- &=& v\grad\left[\rho^-\left(1-\frac{\rho}{\rho_m}\right)\right]+\frac{\alpha\,\rho^+}{2}-\frac{\alpha\,\rho^-}{2}.
\label{eq:ContMF}
\end{eqnarray}
We expand around a flat profile and Fourier transform. We take $\rho^\pm(x)=\rho_0/2+\sum_q\delta_q^\pm\exp(i\,q\,x)$ and arrive at 
\begin{eqnarray}
  \dot{\delta}_q^+ &=& -v\,\delta_q^+\,i\,q\,\left(1-\frac{\rho_0}{\rho_m}\right)+v\,i\,q\,\frac{\rho_0}{2\,\rho_m}\left(\delta_q^++\delta_q^-\right)-\frac{\alpha}{2}\left(\delta_q^+-\delta_q^-\right) \\
   \dot{\delta}_q^- &=& v\,\delta_q^-\,i\,q\,\left(1-\frac{\rho_0}{\rho_m}\right)-v\,i\,q\,\frac{\rho_0}{2\,\rho_m}\left(\delta_q^++\delta_q^-\right)+\frac{\alpha}{2}\left(\delta_q^+-\delta_q^-\right). 
\label{eq:ContDeltaQ}
\end{eqnarray}
We can re-write these two equations in matrix form as
\begin{equation}
\fl \dot{\vec{\delta_q}} = \bpmatrix -v\,i\,q\left(1-\frac{3\,\rho_0}{2\,\rho_m}\right)-\frac{\alpha}{2} & v\,i\,q\frac{\rho_0}{2\,\rho_m}+\frac{\alpha}{2} \\
                         -v\,i\,q\frac{\rho_0}{2\,\rho_m}+\frac{\alpha}{2} & v\,i\,q\left(1-\frac{3\,\rho_0}{2\,\rho_m}\right)-\frac{\alpha}{2}
                        \epmatrix \vec{\delta_q};\qquad \vec{\delta_q}=\bpmatrix  \delta_q^+ \\ \delta_q^-\epmatrix.
\end{equation}
The eigenvalues of this matrix, which will then tell us whether the flat profile is stable or unstable to small perturbations, are
\begin{equation}
 \lambda_\pm(q) = -\frac{\alpha}{2}\pm\left(\frac{\alpha^2}{4}+v^2\,q^2\left(1-\frac{\rho_0}{\rho_m}\right)\left(\frac{2\,\rho_0}{\rho_m}-1\right)\right)^{1/2}.
 \label{eq:ContStabilityEigenvalues}
\end{equation}
It is clear that one of these eigenvalues will always be negative while the other is negative for $\rho_0<\rho_m/2$ and positive for $\rho_0>\rho_m/2$. Hence a homogeneously flat profile is stable when the average total density is less than half the maximum density and unstable above that, with no dependency on run length. This corresponds precisely with the stability predicted by the free energy derived in section~\ref{sec:FiniteRange}. That stability analysis was derived from a free energy which considered only the total density and was itself calculated only after assuming a diffusive scaling. That the diffusive scaling does not alter the criterion for instability justifies taking that limit and implies that the discrepancy between our lattice simulations and continuum free energy arises from another factor.

We turn, then, to consider the dynamic stability of the lattice dynamics directly. Beginning with the mean field equations for the anisotropic partial exclusion process, 
\begin{eqnarray}
  \dot{n}^+_i &=& d\,n_{i-1}^+\left(1-\frac{n_i}{n_{m}}\right)-d\,n_{i}^+\left(1-\frac{n_{i+1}}{n_{m}}\right)-\frac{\alpha\,n_i^+}{2}+\frac{\alpha\,n_i^-}{2} \\
  \dot{n}^-_i &=&  d\,n_{i+1}^-\left(1-\frac{n_i}{n_{m}}\right)-d\,n_{i}^-\left(1-\frac{n_{i-1}}{n_{m}}\right)+\frac{\alpha\,n_i^+}{2}-\frac{\alpha\,n_i^-}{2}
\label{eq:LatticeMF}
\end{eqnarray}
we expand around a flat profile, taking $n_k^\pm=n_0+\sum_q\delta_q^\pm\exp(i\,q\,k)$. After some algebra, detailed in~\ref{app:Stability}, we arrive at a condition for there to exist positive eigenvalues, i.e. for a flat profile to be unstable to small perturbations. Specifically we see instability whenever the run length $r=d/\alpha$ satisfies the following inequality
\begin{equation}
 r > \frac{1}{2\,\left(1-\frac{n_0}{n_{m}}\right)\left(\frac{2\,n_0}{n_{m}}-1\right)}.
\end{equation}

\begin{figure}
 \centering
\includegraphics[width=7cm]{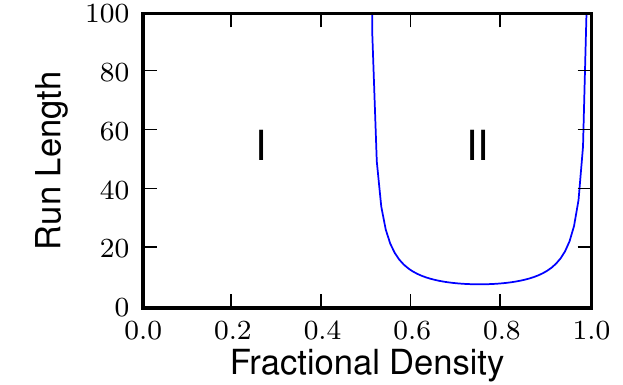}
\caption{A flat profile is stable when in region I, but unstable in region II, for a system with exclusion and homogeneous, isotropic jump and tumble rates. The $x$-axis is the fractional density, i.e. $n/n_{m}$.}
\label{fig:LatticeStability}
\end{figure}

Graphing this we can see that a flat profile will be stable in region I of figure~\ref{fig:LatticeStability} and unstable in region II. We see that for short run lengths the range of densities in which the system will spinodally decompose is restricted and at run lengths below $4$ sites there is no separation at all. Conversely, in the limit that the run length tends to infinity, i.e. where we effectively have two totally asymmetric partial exclusion processes on the same lattice, the system is unstable for any density between $\frac{n_0}{n_{m}}=0.5$ and $\frac{n_0}{n_{m}}=1$. This instability is in accordance with that seen in our simulations.

In our simulations we found that when we replaced the anisotropic kernel in the interaction terms with an isotropic one we recovered the density profiles predicted by the continuous theory. We can also analyse the effect of an isotropic density kernel on the dynamic stability.

Consider now the microscopic mean field equations 
\begin{eqnarray}
\fl  \dot{n}^+ &=& dn_{i-1}^+\Bigg(1-\frac{1}{n_{m}}\sum_jK_j^+n_{i+j-1}\Bigg)-dn_{i}^+\Bigg(1-\frac{1}{n_{m}}\sum_jK_j^+n_{i+j}\Bigg)-\frac{\alpha n_i^+}{2}+\frac{\alpha n_i^-}{2} \\
\fl  \dot{n}^- &=&  dn_{i+1}^-\Bigg(1-\frac{1}{n_{m}}\sum_jK_j^-n_{i+j+1}\Bigg)-dn_{i}^-\Bigg(1-\frac{1}{n_{m}}\sum_jK_j^-n_{i+j}\Bigg)+\frac{\alpha n_i^+}{2}-\frac{\alpha n_i^-}{2}.
\label{eq:GeneralLatticeMF}
\end{eqnarray}
In general $K_j^\pm$ can take any values, we enforce only that they are both normalised, i.e. that $\sum_j K_j^\pm=1$. Relaxing this constraint would effectively re-normalise the maximum density. This more general interaction reduces to the simple exclusion case if we take $K^\pm_j=\delta_{\pm1,j}$ and to the zero-range case if we take $K^\pm_j=\delta_{0,j}$. 

For an isotropic kernel, where $K^\pm_i=K_i$, when we expand around a flat profile we find that there exists a $q$ such that $\lambda_+(q)$ is greater than $0$ if and only if $n_{m}>n_0>0.5\, n_{m}$, see~\ref{app:Stability} for details, which matches the condition we derived from our continuum free energy. Thus for an isotropic interaction kernel we recover the continuum stability result while an anisotropic kernel will, in general, not produce the same result. It thus seems the error comes from the continuum limit itself, which is not valid for anisotropic kernels. While the stability analysis accounts qualitatively for the difference between isotropic and anisotropic kernels a theory which quantitatively accounts for the differences in coexistence densities at steady state remains to be constructed.

\subsection{Finite Range Interactions in 2D}
\label{sec:TwoDimensions}
\begin{figure}
 \centering
  \begin{minipage}{0.48\textwidth}
  \centering
  \includegraphics[width=7cm]{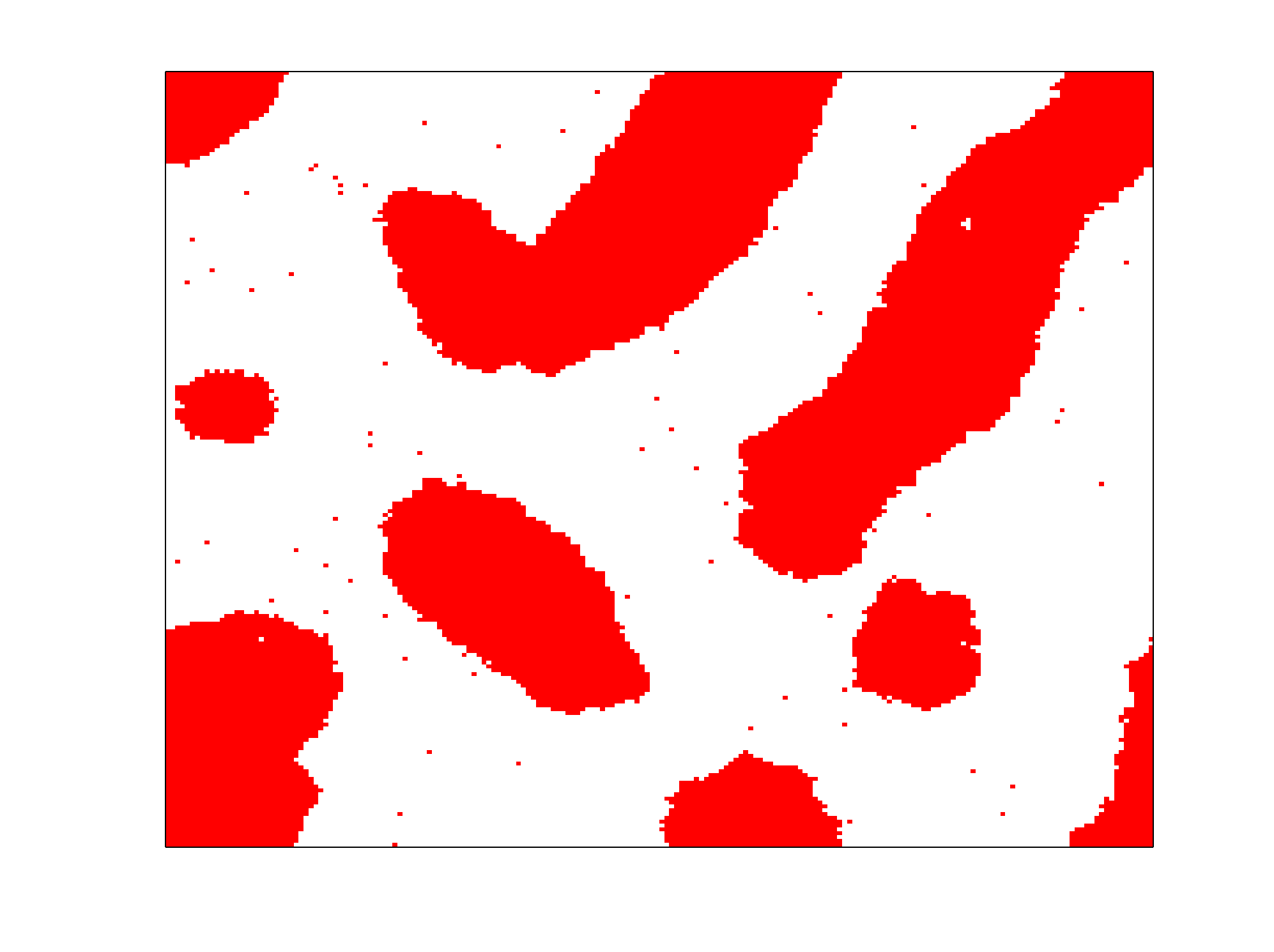}
 \end{minipage}
 \begin{minipage}{0.48\textwidth}
  \centering
  \includegraphics[width=7cm]{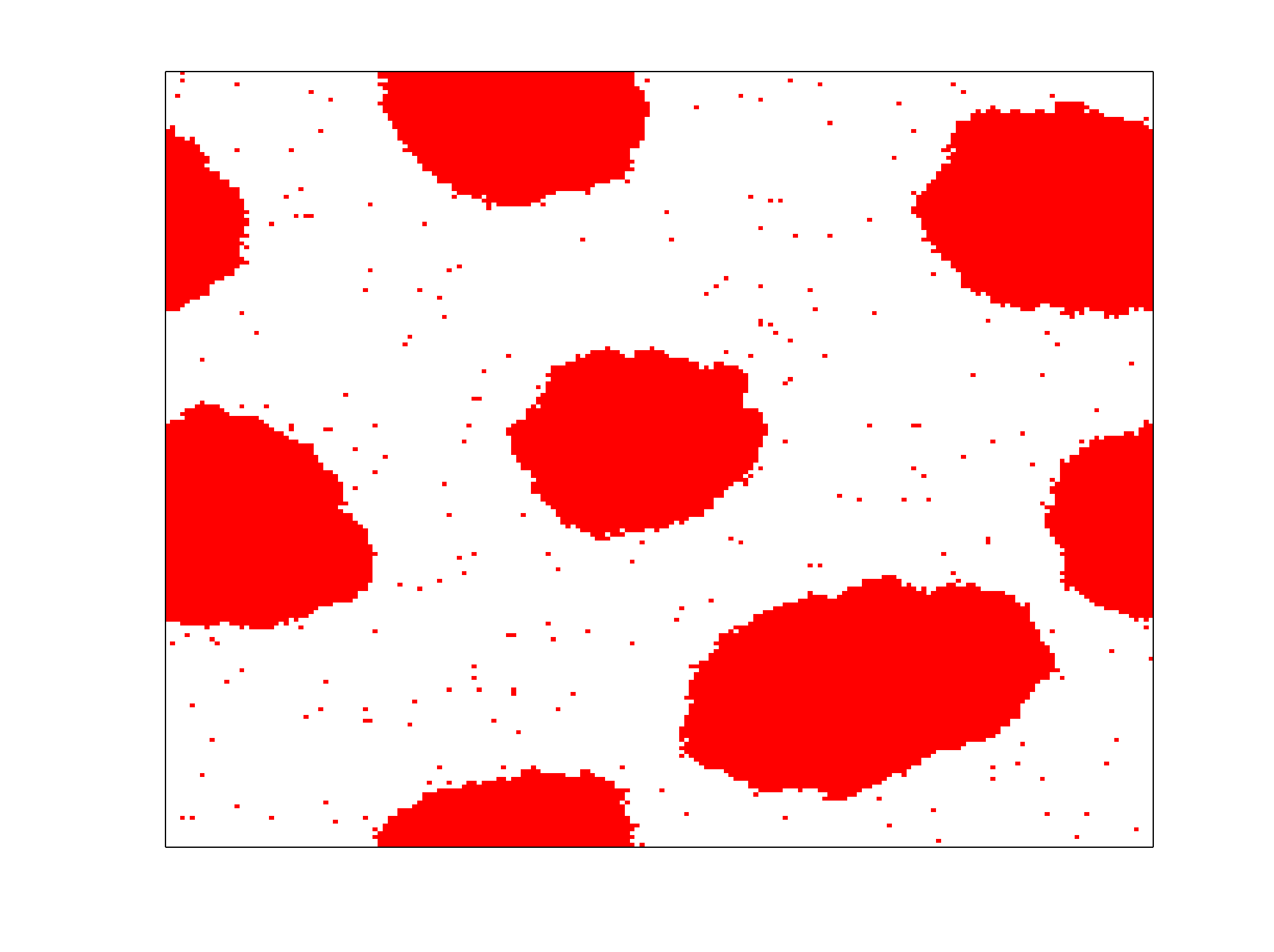}
 \end{minipage}
 \caption{Snapshots of the two dimensional system using the partial asymmetric exclusion and parameters $n_{m}=20$, $d_0=10$ and $\alpha=1$. Sites with a density greater than $0.6n_{m}$ are marked in red, while those with a lower density are left blank. {\bf Left}: allowing only nearest neighbour hops and no diagonal movement. {\bf Right}: allowing for diagonal hops. Both simulations performed using $400,000$ particles and recorded at $t=2500$.}
 \label{fig:2DSnapshots}
\end{figure}

Most cases of physical interest require a model in more than one dimension; it is therefore natural to extend our analysis to higher dimensions. This can be done on lattice relatively easily. On a square lattice in two dimensions we allow the particles to jump between nearest neighbours. We therefore consider $4$ types of particle now instead of $2$. We find that in two dimensions, the behaviour of the run-and-tumble crowding model is qualitatively the same as in one. The system separates into regions of high and low density, where those co-existent densities are given by the same free energy as in one dimension. The field theoretic approach developed in section~\ref{sec:FHIB} indeed generalizes straightforwardly to higher dimensions and yields the same fluctuating hydrodynamics. We find, however, that allowing only nearest neighbour hopping results in an unrealistic surface tension because of the anistropy of the lattice~\cite{Krapivsky2004,Funaki1997}; the regions of high and low density form elongated, and thus anisotropic, domains, see figure~\ref{fig:2DSnapshots} (left).

To correct this unphysical characteristic of our model we extend our dynamics to allow next to nearest hopping along diagonal directions and consider $8$ species of particles, where the jump rates in the diagonal directions are scaled by a factor of $\sqrt{2}$. The droplets then coarsen into more realistic curved domains, see figure~\ref{fig:2DSnapshots} (right). 

 The stability conditions remain the same as do the coexistence densities and the discrepancy between isotropic and anisotropic kernels. We examined simulations with both the anisotropic partial exclusion kernel, and the isotropic, Gaussian kernel on lattices of $500$x$500$ sites for densities above the spinodal point. The systems were seen to separate into droplets of higher and lower density which then coarsened into discrete, contiguous domains, see figure~\ref{fig:2DSnapshots}.

We measured the coarsening of these domains and found them to scale as $t^{1/3}$, as we would expect for conserved model B type dynamics~\cite{Bray1994}. An approximate measurement of the size of the domains was calculated by randomly sampling the system and measuring the horizontal and vertical size of the encountered droplet at that point. Mathematically, we define
\begin{eqnarray}
  L_x(i,j)&=&\max\left\lbrace k\in\mathbb{N}: \abs{n_{i,j}-n_{i+m,j}}<n^\ast,\, \forall m \in \left[0,k\right]\right\rbrace \nonumber \\
&+&\max\left\lbrace k\in\mathbb{N}: \abs{n_{i,j}-n_{i-m,j}}<n^\ast,\, \forall m \in \left[0,k\right]\right\rbrace,
\end{eqnarray}
where $n^\ast$ is an arbitrary cutoff to distinguish the two domains but ignore random fluctuations. Computations were run with a number of choices for $n^\ast$ and the particular choice of cutoff was found to have no significant effect on the results. We calculate the vertical size in an analogous fashion and average the lengths over a large number of points on the lattice. Though this does not give an exact measure of the droplet size it is sufficient to show the scaling of the domain size with time whilst being quick to calculate numerically. 

\begin{figure}
 \centering
\begin{minipage}{0.48\textwidth}
\centering
 \includegraphics[width=7cm]{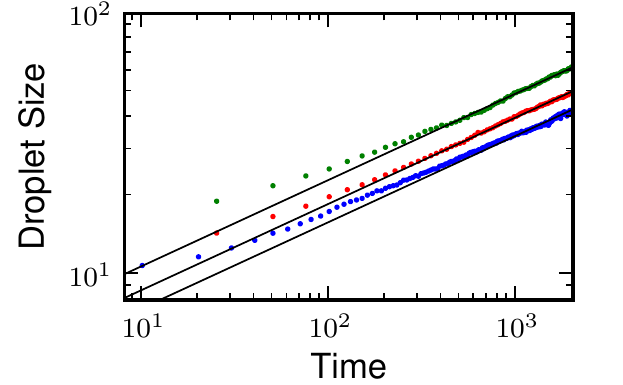}
\end{minipage}
\begin{minipage}{0.48\textwidth}
\centering
 \includegraphics[width=7cm]{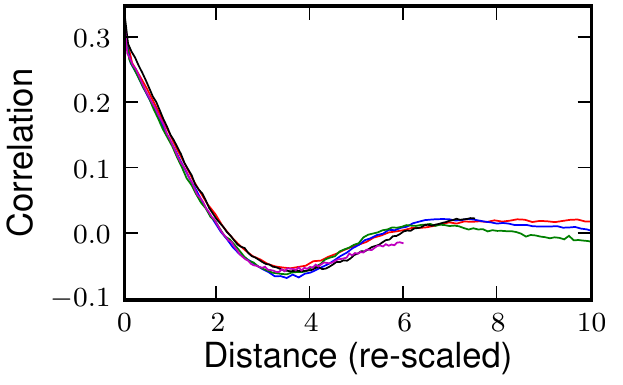}
\end{minipage}
\caption{{\bf Left}: The average diameter of domains over time for an anisotropic kernel (green), Gaussian kernel with $k=1$ (red) and $k=2$ (blue). The solid lines show asymptotic behaviour with an exponent of $1/3$. {\bf Right}: The correlation function $C(j,t)$ for the anisotropic kernel simulations. The x-axis has been rescaled $x\rightarrow x/t^{0.33}$ so that data computed at $t=250$ (red), $t=500$ (blue), $t=1000$ (green), $t=2500$ (black) and $t=5000$ (magenta) superimpose. Both figures derived from data for $240,000$ particles and with parameters $n_m=10$, $\alpha=1$ and $d_0=10$ on a lattice of $200\times200$ sites.}
\label{fig:2DCorrelations}
\end{figure}

Using this procedure we determine that the domains increase in size with an exponent of approximately $1/3$, i.e. $\avg{L_x}(t^\prime)=(t^\prime/t)^{1/3} \avg{L_x}(t)$.  The two-point, connected, equal time correlation function $C(j,t)=\avg{n_i(t)\,n_{i+j}(t)}-\avg{n_i(t)}\avg{n_{i+j}(t)}$ was also calculated numerically from the data and fit reasonably with a re-scaling $C(x,t)=C(x/a^{1/3},t\,a)$. The data for both these measurements can be seen in figure~\ref{fig:2DCorrelations}. 

As can be seen from the figures the choice of kernel does not change the coarsening exponent, only the relative speed of coarsening, with interactions over a larger number of sites taking longer to reach a steady state than those with shorter ranges.

\section{Conclusion}
In this work we have presented several classes of lattice models based on the run-and-tumble dynamics of certain species of bacteria, notably \emph{Escherichia coli}. We calculated the exact steady state probability distributions for both inhomogeneous, anisotropic, non-interacting and zero-range interaction models. For more general types of interaction we used a field theoretic approach to derive the continuum fluctuating hydrodynamics and from there derived a mapping to a free energy like functional describing the steady state profile for a crowding interaction. We analysed the linear stability of both the continuum and lattice microscopic mean field equations and isolated a condition on the coarse graining we employed in our interaction terms for the zero-range free energy to work as a mean field theory.

Our work builds on earlier treatments of run-and-tumble bacteria where interactions between bacteria were not included~\cite{Schnitzer1990,Schnitzer1993}. It provides a lattice counterpart to prior continuum approaches and qualitatively reproduces results obtained off lattice, where a similar, though not exactly equivalent, density dependence was considered~\cite{Tailleur2008}. Our approach on lattice provides a microscopic justification for manner in which this density dependence manifests; previously it was added in an ad-hoc manner after the diffusive approximation had been taken where in this work the dependence is intrinsic from the microscopic definition of the dynamics. That we produce qualitatively similar results justifies the way in which the dependence was inserted there. Our work also reveals a condition on the way in which the coarse graining in this density dependence must be performed for free energy mapping to work; it must be taken isotropically, so that particles moving in any direction feel the same effective local density.

This work provides a means to simulate microscopic run-and-tumble dynamics efficiently, which is particularly important in two or more dimensions where microscopic simulations off lattice are very computationally expensive. It illustrates potential hazards in comparing lattice simulations and continuum theoretical predictions and offers some insight into how to avoid those problems by carefully choosing how to implement non-local interactions in the lattice dynamics. Note that early studies of off-lattice run-and-tumble dynamics actually had to discretise space to do simulation for position dependent swimming speed in order to compare with their theoretical predictions~\cite{Schnitzer1990}. We hope the techniques we presented in this paper will develop the use of lattice simulations in the bacterial context much further, for example, to study pattern formation.

We hope that the lattice approach will provide new exact results for run-and-tumble dynamics in higher dimensions, generalizing previous studies to more general external potentials for sedimentation and trapping~\cite{Tailleur2009}.

Last, there are some open question regarding the coexistence densities for interacting bacteria. Although our fluctuating hydrodynamics correctly predicts the existence of phase separation, the coexistence densities are accurately predicted only for isotropic kernels and large lattice occupancies. Quantitative predictions beyond this case are yet to be derived. More generally, explicit two body interactions and more general non-linear interaction kernels have not been looked at yet. These are a few of the outstanding questions among the many concerning this new and interesting class of models.	

%%%%%%%%%%%%%%%%%%%%%%%%%%%%%%%%%%%%%%%%%%%%%%%%%%%%%%%%%%%%%%%%%%%

\section*{Acknowledgements}
We thank V. Lecomte, F. van Wijland and R.K.P. Zia for fruitful discussions and
acknowledge funding from {\em the Carnegie Trust for the Universities
of Scotland} (A.G.T.), EPRSC EP/E030173 (M.E.C.) and EP/H027254 (J.T.) and RCUK
(R.A.B.). M.E.C. holds a Royal Society Research Professorship.

%%%%%%%%%%%%%%%%%%%%%%%%%%%%%%%%%%%%%%%%%%%%%%%%%%%%%%%%%%%%%%%%%%%

\section*{References}

\bibliographystyle{unsrt}
\bibliography{references}

\pagebreak
\appendix
  \renewcommand{\theequation}{A-\arabic{equation}}
\section{Hydrodynamic Limit and Scaling of Fields in the Action}
\label{app:Scaling}
The microscopic action for the non-interacting, homogeneous and isotropic model can be written as
\begin{eqnarray}
\fl    S&=&-\int_0^T \sum_i\bigg[ \hat \rho_i \dot \rho_i + \frac 1 d \hat J_i
    \dot J_i + \frac d 2 \rho_i \left(e^{-(\hat
    \rho_{i+1}-\hat \rho_i + \hat J_{i+1}-\hat J_i)}+
e^{\hat \rho_{i+1}-\hat \rho_i - (\hat J_{i+1}-\hat
      J_i)}-2\right)\nonumber\\
\fl     &+&\frac {J_i}2 \left(e^{-(\hat
    \rho_{i+1}-\hat \rho_i + \hat J_{i+1}-\hat J_i)} -
    e^{\hat \rho_{i+1}-\hat \rho_i - (\hat J_{i+1}-\hat
    J_i)}\right)\nonumber\\
\fl     &+&\frac {d}2(\rho_{i+1}-\rho_i-\frac{J_{i+1}-J_i}d)
    \left( e^{ \rho_{i+1}-\hat \rho_i - (\hat
    J_{i+1}-\hat J_i)}-1\right)\nonumber\\
\fl     &+&\frac {\alpha \rho_i}4 \left(e^{2 \hat J_i}+e^{-2 \hat J_i}-2\right)+\frac {\alpha J_i}{4d} \left(e^{2 \hat J_i}-e^{-2 \hat J_i}\right)\bigg].
\end{eqnarray}
The continuous limit can be taken by explicitly introducing the
lattice spacing $a$ and making the substitutions
\begin{eqnarray}
    &\rho_i\to a\rho(x);\quad \hat\rho_i\to\hat\rho(x);\quad d\to v a^{-1};\quad\sum_i
  \to \int_0^{\ell=L a} \dx a^{-1};\nonumber\\
  &\quad J_i\to J(x);\quad \hat J_i\to \hat J(x);\quad \grad_i \to a \grad +\frac 1 2  a^2 \Delta
\end{eqnarray}
where $\grad_i$ is the discrete gradient, e.g. $\grad_i
\rho_i=\rho_{i+1}-\rho_i$. This overall substitution and the Taylor
expansion of the action then gives
\begin{eqnarray}
\fl   S&=&-\int_0^T \dt \int_0^\ell \dx\;\bigg[ \hat \rho \dot \rho +v^{-1} \hat J \dot J - v \rho \grad \hat J - J \grad \hat \rho +\frac{\alpha \rho}4 \left(e^{2 \hat J}+e^{-2 \hat J}-2\right)\nonumber \\
\fl&+&\frac{\alpha J}{4 v} \left(e^{2\hat J}-e^{-2\hat J}\right)\bigg] +a S_1,
\label{eq:ActionPreScaling}
\end{eqnarray}
where the neglected term $S_1$ is given by
\begin{eqnarray}
\fl  S_1&=&-\int_0^T\dt\int_0^\ell  \dx\;\bigg[ v \rho \big[-\frac 1 2 \Delta \hat J+(\grad \hat\rho)^2/2+(\grad
  \hat J)^2/2\big]+J\big[-\Delta \hat \rho/2+\grad \hat\rho\grad \hat
    J\big]\nonumber \\
\fl &+&v/2\big(\grad \rho-\grad J/v\big)\big(\grad \hat \rho-\grad \hat J\big)\bigg]+{\cal O}(a).
\end{eqnarray}
To calculate the correct manner in which to rescale our fields let us begin by considering a system $\ell$ times larger and rescaling $t\rightarrow t \ell^\alpha$, $x\rightarrow x \ell$, $\rho\rightarrow \rho/\ell$ so that the action is given by
\begin{eqnarray}
\fl S &=& - \int_0^{T/\ell^\alpha}\!\! \dt \int_0^1 \dx\; \bigg[\hat \rho \dot \rho + \ell \frac{\hat J \dot J}{v} - \ell^{\alpha-1} v \rho \grad \hat J - \ell^\alpha J \grad \hat \rho - \ell^\alpha \frac{\alpha}{4} \rho \left(e^{2 \hat J}+e^{-2 \hat J}-2\right)\nonumber\\
\fl &-& \ell^{\alpha+1} \frac{\alpha}{4 v} J \left(e^{-2 \hat J} - e^{2 \hat J}\right)\bigg] 
\end{eqnarray}
For the $\hat J$ terms to not blow up we need to have $\hat J$ small. We therefore expand the exponentials to give
\begin{equation}
\fl S = - \int_0^{T/\ell^\alpha}\!\!  \dt \int_0^1 \dx\; \bigg[\hat \rho \dot \rho + \ell \frac{\hat J \dot J}{v} - \ell^{\alpha-1} v \rho \grad \hat J - \ell^\alpha J \grad \hat \rho - \ell^\alpha \alpha \rho \hat J^2 + \ell^{\alpha+1} \frac{\alpha}{v} J \hat J\bigg].
\end{equation}
If we explicitly take $\hat J$ to scale as $\hat J\rightarrow \hat J \ell^{-\beta}$ and  $J\rightarrow J \ell^{-\delta}$ we get
\begin{eqnarray}
S &=& - \int_0^{T/\ell^\alpha}\!\!  \dt \int_0^1 \dx\;\bigg[ \hat \rho \dot \rho + \ell^{1-\beta-\delta} \frac{\hat J \dot J}{v} - \ell^{\alpha-1-\beta} v \rho \grad \hat J \nonumber \\
&-& \ell^{\alpha-\delta} J \grad \hat \rho - \ell^{\alpha-2 \beta} \alpha \rho \hat J^2 + \ell^{\alpha-\beta-\delta+1} \frac{\alpha}{v} J \hat J\bigg].
\end{eqnarray}
Now, we need the coefficient of each term to be of order 1 or smaller so that no terms blow up so
\begin{equation}
\fl 1-\beta-\delta\leq0;\qquad \alpha-1-\beta\leq0;\qquad \alpha-\delta\leq 0;\qquad \alpha-2 \beta\leq 0;\qquad 1+\alpha-\beta-\delta\leq0. 
\end{equation}
 However, as we do not want to simply be left with $\dot \rho = 0$ we need $\alpha-\delta = 0$ and as we also want to retain a noise, which corresponds to the $\hat J^2$ term, we also require that $\alpha-2 \beta = 0$. That leaves our action 
\begin{equation}
\fl S = - \int_0^{T/\ell^\alpha}\!\!  \dt \int_0^1 \dx\;\bigg[ \hat \rho \dot \rho + \ell^{1-3 \beta} \frac{\hat J \dot J}{v} - \ell^{\beta-1} v \rho \grad \hat J - J \grad \hat \rho - \alpha \rho \hat J^2 + \ell^{1-\beta} \frac{\alpha}{v} J \hat J\bigg].
\end{equation}
Which tells us that $\beta\leq 1$, $\beta\geq 1/3$ and $\beta \geq 1$ which imply that $\beta=1$ and hence $\alpha=2$ and $\delta=2$. Injecting these scalings back into the action gives
\begin{equation}
\fl  S=-\int_0^{T/\ell^2}\dt \int_0^1  \dx\; \bigg[\hat \rho \dot \rho +\ell^{-2} v^{-1} \hat J \dot J -
  v \rho \grad \hat J - J \grad \hat \rho +\alpha \rho \hat
  J^2+\frac{\alpha J \hat J}{ v}\bigg]
\end{equation}
where the macroscopic observation time $\tau=T/\ell^2$ is supposed to be of order 1. The term in $\hat J \dot J$ is thus irrelevant and we 
can check that the hydrodynamic action
\begin{equation}
  S_0=-\int_0^\tau \dt\int_0^1 \dx\; \bigg[\hat \rho \dot \rho - v \rho \grad \hat J - J \grad \hat \rho +\alpha \rho \hat  J^2+\frac{\alpha J \hat J}{ v}\bigg]
\end{equation}
is invariant under further diffusive scaling. Note that the scaling of the fields considered here is arbitrary and we could {\emph{choose}} to look at currents $J$, $\hat J$ larger than $1/\ell^2$, $1/\ell$. This would correspond to trajectories whose probabilities are smaller than $\exp(-\ell)$, which are not correctly described by fluctuating hydrodynamics and large deviations. One can also check
that under this rescaling, the action $S_1$ stays of order 1 and $a
S_1$ is thus, indeed, negligible.

For one isolated bacterium the run-and-tumble dynamics is a variant of a random walk and is known to be diffusive at large scales~\cite{Schnitzer1993,Tailleur2008}. It is therefore not surprising that we find $\alpha=2$. In the presence of interactions (as in section~\ref{sec:stocheqI}), a uniform density profile of bacteria will continue to exhibit diffusive behaviour; the interactions will simply rescale the diffusivity. If interactions cause the profile to become unstable, however, the model can nevertheless give rise to length scales which grow in a non-diffusive manner, as for example for coarsening (see section~\ref{sec:TwoDimensions}).

\section{Stability Analyses}
\label{app:Stability}
Beginning with the mean field equations for the partial exclusion-like interaction,
\begin{eqnarray}
  \dot{n}^+ &=& d\,n_{i-1}^+\left(1-\frac{n_i}{n_{m}}\right)-d\,n_{i}^+\left(1-\frac{n_{i+1}}{n_{m}}\right)-\frac{\alpha\,n_i^+}{2}+\frac{\alpha\,n_i^-}{2} \\
  \dot{n}^- &=&  d\,n_{i+1}^-\left(1-\frac{n_i}{n_{m}}\right)-d\,n_{i}^-\left(1-\frac{n_{i-1}}{n_{m}}\right)+\frac{\alpha\,n_i^+}{2}-\frac{\alpha\,n_i^-}{2}
\label{eq:LatticeMFApp}
\end{eqnarray}
we can expand around a flat profile and take $n_k^\pm=n_0+\sum_q\delta_q^\pm\exp(i\,q\,k)$ to investigate the linear stability. In matrix form the resulting equations can be written as
\begin{equation}
\fl \dot{\vec{\delta_q}}\! =\! \bpmatrix \!\!\!\!d\left(\!1-\frac{n_0}{n_{m}}\right)\!\left(e^{-iq}-1\right)+\frac{dn_0}{2n_{m}}\!\left(e^{iq}-1\right)-\frac{\alpha}{2} &\!\!\!\! \frac{dn_0}{2n_{m}}\left(e^{iq}-1\right)+\frac{\alpha}{2} \\
                       \!\!\!\!\!\!\!\!\!\!\!\!\!\!\!\!\!\!\!\!\!\!\!\!  \frac{dn_0}{2n_{m}}\left(e^{-iq}-1\right)+\frac{\alpha}{2} & \!\!\!\!\!\!\!\!\!\!\!\!\!\!\!\!\!\!\!\!\!\!\!\!\!\!\!\!d\left(\!1-\frac{n_0}{n_{m}}\right)\!\left(e^{iq}-1\right)+\frac{dn_0}{2n_{m}}\!\left(e^{-iq}-1\right)-\frac{\alpha}{2}
                        \!\!\epmatrix\! \vec{\delta_q},
\end{equation}
where $\vec{\delta_q}=(\delta_q^+,\delta_q^-)$ as before. Defining the run length as the ratio $d/\alpha=r$, we can write the eigenvalues of this matrix as
\begin{eqnarray}
\fl \lambda_\pm(q) &=& \alpha\Bigg(-\frac{1}{2} + r\left(1-\frac{n_0}{2\,n_{m}}\right)\left(\cos(q)-1\right) \nonumber \\
\fl&\pm&\left[-r^2\left(1-\frac{3\,n_0}{2\,n_{m}}\right)^2\sin^2(q)+\frac{1}{4}+\left(\frac{r^2\,n_0^2}{2\,n_{m}^2}-\frac{r\,n_0}{2\,n_{m}}\right)\left(1-\cos(q)\right)\right]^{\frac 1 2}\Bigg).
\end{eqnarray}
Again, one eigenvalue is always negative while one can be positive or negative. In this case, however the condition for stability is no longer independent of $q$ and $r$. In particular, we find that $\lambda_+>0$ when 
\begin{equation}
\fl -2\,r^2\,\left(\frac{n_0}{n_{m}}\right)^2-2\,r^2+4\,r^2\,\left(\frac{n_0}{n_{m}}\right)-r+\cos(q)\,(2\,r^2\,\left(\frac{n_0}{n_{m}}\right)-2\,r^2\,\left(\frac{n_0}{n_{m}}\right)^2)>0.
\label{eq:LatticeStabilityQCond}
\end{equation}
Further, if we want to know only under what conditions on $r$ and $\frac{n_0}{n_{m}}$ there exist any positive eigenvalues, we can consider the simpler condition
\begin{equation}
\fl -2\,r^2\,\left(\frac{n_0}{n_{m}}\right)^2-2\,r^2+4\,r^2\,\left(\frac{n_0}{n_{m}}\right)-r+2\,r^2\,\left(\frac{n_0}{n_{m}}\right)-2\,r^2\,\left(\frac{n_0}{n_{m}}\right)^2>0,
\label{eq:LatticeStabilityMaxCond}
\end{equation}
which implies that
\begin{equation}
 r > \frac{1}{2\,\left(1-\frac{n_0}{n_{m}}\right)\left(\frac{2\,n_0}{n_{m}}-1\right)}.
\end{equation}

For the more general case where the microscopic mean field equations are given by
\begin{eqnarray}
  \dot{n}^+ &=& d\,n_{i-1}^+\Bigg(1-\frac{1}{n_{m}}\sum_j\,K_j^+\,n_{i+j-1}\Bigg)-d\,n_{i}^+\Bigg(1-\frac{1}{n_{m}}\sum_j\,K_j^+\,n_{i+j}\Bigg)\nonumber\\
&-&\frac{\alpha\,n_i^+}{2}+\frac{\alpha\,n_i^-}{2} \\
  \dot{n}^- &=&  d\,n_{i+1}^-\Bigg(1-\frac{1}{n_{m}}\sum_j\,K_j^-\,n_{i+j+1}\Bigg)-d\,n_{i}^-\Bigg(1-\frac{1}{n_{m}}\sum_j\,K_j^-\,n_{i+j}\Bigg) \nonumber \\
&+&\frac{\alpha\,n_i^+}{2}-\frac{\alpha\,n_i^-}{2},
\label{eq:GeneralLatticeMFApp}
\end{eqnarray}
we can perform a similar analysis. Once again we expand around a flat profile, this time to obtain
\begin{eqnarray}
\fl  \dot{\delta}_q^+ &= & d\,\delta_q^+\,e^{-i\,q}\Bigg(1-\frac{1}{n_{m}}\sum_j\,K_j^+\,n_0\Bigg)-d\,\delta_q^+\Bigg(1-\frac{1}{n_{m}}\sum_j\,K^+_j\,n_0\Bigg) -\frac{\alpha}{2}\left(\delta_q^+-\delta_q^-\right)\nonumber\\
\fl& -&\frac{d\,n_0}{2\,n_{m}}\left(\delta_q^++\delta_q^-\right)\sum_j\,K^+_j\,e^{i(j-1)q} +\frac{d\,n_0}{2\,n_{m}}\left(\delta_q^++\delta_q^-\right)\sum_j\,K^+_j\,e^{ijq} \\
\fl \dot{\delta}_q^- &= & d\,\delta_q^-\,e^{i\,q}\Bigg(1-\frac{1}{n_{m}}\sum_j\,K_j^-\,n_0\Bigg)-d\,\delta_q^-\Bigg(1-\frac{1}{n_{m}}\sum_j\,K^-_j\,n_0\Bigg) +\frac{\alpha}{2}\left(\delta_q^+-\delta_q^-\right)\nonumber\\
\fl& -&\frac{d\,n_0}{2\,n_{m}}\left(\delta_q^++\delta_q^-\right)\sum_j\,K^-_j\,e^{i(j+1)q} +\frac{d\,n_0}{2\,n_{m}}\left(\delta_q^++\delta_q^-\right)\sum_j\,K^-_j\,e^{ijq}.
\label{eq:GeneralStabilityDeltaQ}
\end{eqnarray}
Now, we can define a new function $\kappa^\pm(q)=\sum_j K_j^\pm \exp(i\,j\,q)$. With this definition, and the normalisation of $K^\pm_j$, we can simplify equations~\eqref{eq:GeneralStabilityDeltaQ}.
\begin{equation}
\fl \dot{\vec{\delta_q}}\! = \!\bpmatrix \!\!\!d\left(\!1-\frac{n_0}{n_{m}}-\frac{dn_0}{2n_{m}}\kappa^+(q)\!\right)\!\left(e^{-iq}-1\right)-\frac{\alpha}{2} & \!\!-\frac{dn_0}{2n_{m}}\left(e^{-iq}-1\right)\kappa^+(q)+\frac{\alpha}{2} \\
                         \!\!\!\!\!\!\!\!-\frac{dn_0}{2n_{m}}\left(e^{iq}-1\right)\kappa^-(q)+\frac{\alpha}{2} & \!\!\!\!\!\!\!\!\!\!\!\!\!d\left(\!1-\frac{n_0}{n_{m}}-\frac{dn_0}{2n_{m}}\kappa^-(q)\!\right)\!\left(e^{iq}-1\right)-\frac{\alpha}{2}
                        \!\!\epmatrix\! \vec{\delta_q},
\label{eq:GeneralLatticeStabiltyMatrix}
\end{equation}
For isotropic kernels, where $\kappa^\pm(q)=\kappa(q)$, it turns out we can write the eigenvalues in a relatively simple form and, even without knowledge of the specific shape of $K_j$, we can analyse the conditions under which there will exist positive eigenvalues. We start by writing the eigenvalues of the matrix in equation~\eqref{eq:GeneralLatticeStabiltyMatrix} as
\begin{eqnarray}
\fl \lambda_\pm(q)&= &\alpha\Bigg(-\frac{1}{2}+r\left(1-\frac{n_0\left(1+1/2\kappa(q)\right)}{n_{m}}\right)\left(\cos(q)-1\right)\nonumber\\
\fl&\pm&\bigg[-r^2\left(1-\frac{n_0\left(1+1/2\kappa(q)\right)}{n_{m}}\right)^2\left(1-\cos(q)\right)\left(1+\cos(q)\right)\nonumber\\
\fl&+&\frac{1}{4} -\frac{1}{2}\frac{r\,n_0\,\kappa(q)}{n_{m}}\left(\cos(q)-1\right)+\frac{1}{2}\frac{r^2\,n_0^2\,\kappa^2(q)}{n_{m}}^2\left(1-\cos(q)\right)\bigg]^{1/2}\Bigg).
\end{eqnarray}
The larger of these two eigenvalues will be positive if
\begin{equation}
\fl -2\,r^2-r+\frac{r\,n_0}{n_{m}}-\frac{2\,r^2\,n^2_0}{n_{m}^2}+\frac{4\,r^2\,n_0}{n_{m}}+\kappa(q)\left(\frac{2\,r^2\,n_0}{n_{m}}-\frac{2\,r^2\,n_0^2}{n_{m}^2}+\frac{r\,n_0}{n_{m}}\right) > 0.
\end{equation}
Further, as $K(x)$ is always positive, $\kappa(q)$ will have a maximum at $q=0$, where $\kappa(0)=1$, and so we can examine the simpler condition
\begin{equation}
 -2\,r^2-r+\frac{2\,r\,n_0}{n_{m}}-\frac{4\,r^2\,n^2_0}{n_{m}^2}+\frac{6\,r^2\,n_0}{n_{m}} > 0.
\end{equation}
When this inequality is fulfilled we will see instability. Given that $r$ must be positive this means any isotropic density kernel will be unstable in the range $n_0\in[0.5\,n_{m},n_{m}]$. 
\end{document}